\documentclass[%
 aip,
 amsmath,amssymb,
 reprint,%
]{revtex4-1}

\usepackage{graphicx}
\usepackage{dcolumn}
\usepackage{bm}

\usepackage[utf8]{inputenc}
\usepackage[T1]{fontenc}
\usepackage{mathptmx}
\usepackage{etoolbox}

\makeatletter
\def\@email#1#2{%
 \endgroup
 \patchcmd{\titleblock@produce}
  {\frontmatter@RRAPformat}
  {\frontmatter@RRAPformat{\produce@RRAP{*#1\href{mailto:#2}{#2}}}\frontmatter@RRAPformat}
  {}{}
}%
\makeatother
\usepackage{graphicx}
\usepackage{bm}
\usepackage[usenames, dvipsnames]{color}
\usepackage{amsmath}
\usepackage{amssymb}
\usepackage{braket}
\usepackage{changes}
\definecolor{darkblue}{rgb}{0, 0, 0.5}

\usepackage{amsmath}
\usepackage{amssymb}
\usepackage{braket}

\newcommand{\be}{\begin{equation}}
\newcommand{\ee}{\end{equation}}
\newcommand{\ti}[1]{\text{#1}}
\newcommand{\mc}[1]{\mathcal{#1}}

\newcommand{\fig}[1]{Fig.~\ref{#1}}
\newcommand{\app}[1]{Appendix ~\ref{#1}}

\newcommand{\Fig}[1]{Figure~\ref{#1}}

\newcommand{\eq}[1]{Eq.~\eqref{#1}}
\newcommand{\eqs}[2]{Eqs.~\eqref{#1} and \eqref{#2}}
\newcommand{\Eq}[1]{Equation~\eqref{#1}}
\newcommand{\Eqs}[2]{Equations~\eqref{#1} and \eqref{#2}}
\newcommand{\stn}[1]{Sec.~\ref{#1}}

\newcommand{\bea}{\begin{eqnarray}}
\newcommand{\eea}{\end{eqnarray}}
 
\newcommand{\ba}{\begin{array}}
\newcommand{\ea}{\end{array}}

\newcommand{\bl}{\begin{flalign}}
\newcommand{\enl}{\end{flalign}}

\newcommand{\tr}{\text{Tr}}

\begin{document}

\title{General framework for quantifying dissipation pathways in open quantum systems. III. Off-diagonal system-bath couplings}

\author{Ignacio Gustin}
\affiliation{
    Department of Chemistry, University of Rochester, Rochester, New York 14627, USA
    }
  \author{Chang Woo Kim}
\affiliation{
    Department of Chemistry, Chonnam National University, Gwangju 61186, South Korea
   }
\affiliation{
    The Research Institute for Molecular Science, Chonnam National University, Gwangju 61186, South Korea}

\author{Ignacio Franco}
\email{ignacio.franco@rochester.edu}
\affiliation{
    Department of Chemistry, University of Rochester, Rochester, New York 14627, USA
    }
\affiliation{
    Department of Physics, University of Rochester, Rochester, New York 14627, USA
    }
\affiliation{Institute of Optics, University of Rochester, Rochester, New York 14627 USA}

\date{\today}
\begin{abstract}
This paper extends the previously reported theory of dissipation pathways [J. Chem. Phys. 160, 214111
(2024)] to incorporate off-diagonal subsystem-bath coupling, which is often required to model molecular systems where the environment directly influences transitions and couplings between subsystem states. We systematically derive master equations for both population transfer and dissipation into individual bath components, for which we also rigorously prove energy conservation and detailed balance. The approach is based on second-order perturbation theory with respect to the subsystem-bath couplings, whose form is not limited to any specific model. The accuracy of the developed method is tested by applying it to diverse model Hamiltonians involving linearly coupled harmonic oscillator baths and comparing the outcomes against the hierarchical equations of motion (HEOM) method. Overall, our method accurately quantifies the contributions of specific bath components to the overall dissipation while significantly reducing the computational cost compared to numerically exact methods such as HEOM, thus offering a path to examine how vibronic interactions steer non-adiabatic processes in realistic chemical systems.

\end{abstract}
\keywords{Quantum Dynamics, Open Quantum Systems, Dissipation Pathways}

\maketitle

\section{\label{sec:intro}Introduction}

A wide range of quantum chemical phenomena are non-equilibrium processes where the excess energy is dissipated from the central subsystem to the surrounding environment. Naturally, examining this energy flow in detail can provide fundamental insights to understand chemical reactions, material properties, and essential biological processes. For instance, elucidating the major dissipation pathways is crucial for grasping how photosynthetic complexes transfer energy\cite{mirkovic2017light,jang2018delocalized,Cao2020Quantum} and designing physical systems that enhance or suppress dissipation\cite{Kienzler2014,CamposGonzalezAngulo2019,Ng2020,hart2021engineering}. 

While the basic principles of energy transfer are well-established, elucidating the precise pathways of energy flow remains challenging as it amounts to resolving the energy transport within complex molecular environments at a microscopic level. Addressing this challenge requires a method capable of effectively decomposing the overall dissipation into the contributions of individual environmental components. To accomplish this task, it is necessary to fully capture the dynamical information regarding individual vibronic quantum states, which is often computationally prohibitive even with state-of-the-art simulation methods for quantum dynamics. For example, explicit approaches such as the multi-configurational time-dependent Hartree (MCTDH)\cite{beck2000multiconfiguration} method, which accurately track the dynamics via direct wavefunction propagation, become computationally intractable for macroscopic thermal environments. By contrast, quantum master equations (QME) and related techniques\cite{kundu_tight_2022,strathearn2018efficient,varvelo2021formally,Bose2022multisite} can cope with complex chemical environments by focusing on their implicit effect on the dynamics within the subsystem. However, this ability comes at the cost of losing information regarding the quantum states of the environment.


To address this challenge, we recently introduced QME-D\cite{kim2024general1}, a general theoretical framework for quantifying and resolving dissipation pathways in complex quantum systems involving highly structured thermal environments. The theoretical framework utilizes the Nakajima–Zwanzig projection operator technique\cite{nakajima_quantum_1958,zwanzig_ensemble_1960} which is combined with second-order perturbation theory with respect to the coupling between the subsystem states.\cite{kim2024general1,kim2021theory} The framework was proven to be useful in unraveling the detailed dissipation pathways in the realistic model of Fenna-Matthews-Olson photosynthetic complex.\cite{Gustin_dissipation_2025}


Despite the success of the QME-D in studying the quantum dynamics of molecular systems, its applicability is still limited by the assumption that the bath couples only to the diagonal part of the subsystem Hamiltonian matrix. Under such settings, the bath only modulates the energies of the subsystem states and therefore does not directly mediate population transfer. In molecular systems, this is equivalent to the Condon approximation under which couplings between molecular electronic states are unaffected by the nuclei. However, there are various situations where intramolecular vibrations or solvent can actively modulate the electronic couplings to induce non-adiabatic transitions. These considerations motivate us to generalize the previously reported framework for quantifying dissipation pathways to handle both on- and off-diagonal system-bath coupling. As a result, the bath is allowed to directly mediate transitions between system states, which better describes the dynamics occurring in a broad range of quantum transport process involving molecules.

The structure of this paper is as follows: In \stn{subsection:population} through \ref{sec:thermo}, we provide an overview of the theoretical background required to understand the main findings of our work and introduce the extended framework for quantifying dissipation pathways. Subsequently, \stn{subsection:hobath} applies the newly developed approach to specific model Hamiltonians and connect the outcomes with previously established results. In \stn{sec:results-5}, we extensively test the accuracy of our framework against a numerically exact dissipation calculation based on HEOM, while also comparing the performance with QME-D\cite{kim2024general1,kim2024general2} to highlight the utility of the new approach. \stn{sec:conclusions-5} concludes by summarizing the principal findings and discussing conceivable future research directions.

\section{Theory}\label{sec:theory-5}
\subsection{Population transfer}\label{subsection:population}
We take the standard viewpoint for open quantum system dynamics and divide the system Hamiltonian $\hat{H}$ as
\begin{equation}\label{eq:Hfull}
    \hat{H} = \hat{H}_\text{S} + \hat{H}_\text{B} + \hat{H}_\text{SB},
\end{equation}
where $\hat{H}_\text{S}$, $\hat{H}_\text{B}$, and $\hat{H}_\text{SB}$ are the Hamiltonian components for the system, the bath, and the system-bath interaction, respectively. The system Hamiltonian $\hat{H}_\text{S}$ generally takes the form
\begin{equation}\label{eq:Hsub_dia}
    \hat{H}_\text{S} = \sum_{A}\sum_{B} H_{AB} \ket{A}\bra{B},
\end{equation}
where uppercase Roman alphabets are used to label the individual system states, which will be referred to as diabatic basis throughout the rest of this paper. The elements of $\hat{H}_\text{S}$ satisfy $H_{AB} = H_{BA}^*$ due to the Hermicity. 

Having specified the system, we assume that the rest of the Hamiltonian can be split into contributions from independent bath components. This is formally expressed as
\begin{equation}\label{eq:Hbath_int_sum}
    \hat{H}_\text{B} + \hat{H}_\text{SB} = \sum_j \hat{h}_j,
\end{equation}
where $\{ \hat{h}_j \}$ arises from the $j$th bath component and can take a general form of
\begin{equation}\label{eq:Hbathcomp_dia}
    \hat{h}_j = \sum_{A}\sum_{B} (\ket{A} \bra{B} \otimes \hat{v}_{AB}^j).
\end{equation}
The Hermicity requires that the operators in the bath subspace satisfy $\hat{v}_{AB}^j=(\hat{v}_{AB}^j)^\dagger$. The individual bath components only interact through the subsystem and do not directly affect each other, establishing the commutativity between operators with different
$j$'s. We note that \Eq{eq:Hbathcomp_dia} is in contrast to the work presented in Ref. \cite{kim2024general1}, where we only allowed the bath to couple to the system through the diagonal component of $\hat{H}_{\text{S}}$.

We now diagonalize $\hat{H}_\text{S}$ and compute its eigenenergies and eigenstates,
\begin{equation}\label{eq:Hsub_exci}
    \hat{H}_\text{S} = \sum_\alpha E_\alpha \ket{\alpha} \bra{\alpha},
\end{equation}
where each eigenstate $\ket{\alpha}$ is a linear superposition of the diabatic states,
\begin{equation} \label{eq:exc-basis}
    \ket{\alpha} = \sum_{\alpha}c_{\alpha A}\ket{A},
\end{equation}
The basis $\{\ket{\alpha}\}$ is often called the exciton basis, and will be labeled with the Greek alphabet from now on. Recasting \eq{eq:Hbathcomp_dia} using \eq{eq:exc-basis} yields
\begin{equation}\label{eq:Hbathcomp_exci}
    \hat{h}_j = \sum_{\alpha, \beta} ( \ket{\alpha} \bra{\beta} \otimes \hat{v}_{\alpha \beta}^j ),
\end{equation}
where the bath-related operators are transformed as
\begin{equation}\label{eq:v_exci}
    \hat{v}_{\alpha \beta}^j = \sum_{A}\sum_{B} c_{\alpha A}^* c_{\beta B} \hat{v}_{AB}^j.
\end{equation}
It should be noted that there is a freedom of choice for the boundary between the system and the bath, on which the elements of $\hat{H}_\text{S}$ [\eq{eq:Hsub_dia}] depend. Hence, the exciton basis $\{\ket{\alpha}\}$ and the transformation coefficients $\{c_{\alpha A}\} $ are not uniquely determined. We will revisit this point in Sec. \ref{subsection:hobath} where we apply our theory to specific model systems.

The density operator $\hat{\rho}$ for the system evolves according to the Liouville-von Neumann equation $d\hat{\rho}(t)/dt = -i \mc{L} \hat{\rho}(t)/\hbar$, where $\mc{L}$ is the Liouvillian super-operator whose action is defined as $\mc{L} \hat{O} = [\hat{H}, \hat{O}]$ for an arbitrary operator $\hat{O}$. We apply the perturbation theory in the exciton basis by dividing $\hat{H}$ into the diagonal component $\hat{H}_0$ and off-diagonal component $\hat{H}_1$, namely
\begin{subequations}
\begin{equation}
    \hat{H}_0 = \sum_{\alpha} [ \ket{\alpha} \bra{\alpha} \otimes ( E_{\alpha} + \hat{V}_{\alpha \alpha} ) ],
\end{equation}
\begin{equation}
    \hat{H}_1 = \sum_{\alpha} \sum_{\beta \neq \alpha} ( \ket{\alpha} \bra{\beta} \otimes \hat{V}_{\alpha \beta} ) + \ti{H.c.},
\end{equation}
\end{subequations}
and treat $\hat{H}_1$ as the perturbation. In the above, H.c. is the abbreviation for the Hermitian conjugate, and the collective bath operators $\{\hat{V}_{\alpha \beta} \}$ are defined as
\begin{equation}\label{eq:bath_exci_component}
    \hat{V}_{\alpha \beta} = \bra{\alpha} \hat{H}_\text{B} + \hat{H}_\text{SB} \ket{\beta} = \sum_j \hat{v}_{\alpha \beta}^j.
\end{equation}
The Liouvillian is also accordingly divided into $\mc{L} = \mc{L}_0 + \mc{L}_1$, where $\mc{L}_0 \hat{O} = [\hat{H}_0, \hat{O}]$ and $\mc{L}_1 \hat{O} = [\hat{H}_1, \hat{O}]$.

We now apply the projection operator technique\cite{zwanzig_ensemble_1960,nakajima_quantum_1958} to derive the quantum master equation for the evolution of $\hat{\rho}(t)$. We begin by splitting the identity super-operator in the Liouville space into $\mc{I} = \mc{P} + \mc{Q}$ where $\mc{P}$ and $\mc{Q}$ project $\hat{\rho}$ onto the dynamically relevant part $\mc{P}\hat{\rho}$ and the remaining $\mc{Q}\hat{\rho}$, respectively. Because $\mc{P}$ and $\mc{Q}$ are projection operators, they should satisfy $\mc{P}^2 = \mc{P}$ and also $\mc{P}\mc{Q} = \mc{Q}\mc{P} = 0$. At this point, we specify the form of $\mc{P}$ as
\begin{equation}\label{eq:proj_op}
    \mc{P} \hat{\rho} = \sum_\alpha P_\alpha \ket{\alpha} \bra{\alpha} \otimes \hat{R}_\alpha,
\end{equation}
where $P_\alpha = \tr_\ti{b} \bra{\alpha} \hat{\rho} \ket{\alpha}$ is the population of the state $\ket{\alpha}$, $\tr_\ti{b}$ indicates the trace over the bath, and $\hat{R}_\alpha$ is the equilibrium bath density associated with $\hat{V}_{\alpha\alpha}$,
\begin{equation}\label{eq:bathdensity}
    \hat{R}_{\alpha} = \frac{\exp(-\beta \hat{V}_{\alpha\alpha})}{\tr_\ti{b} [\exp(-\beta \hat{V}_{\alpha\alpha})]}.
\end{equation}
The inverse temperature $\beta = 1/k_\ti{B} T$ should not be confused with the exciton index $\beta$, which is only used as a subscript.

At initial time, we assume that the system density is confined in the dynamically relevant part, that is, $\mc{P}\hat{\rho}(0) = \hat{\rho}(0)$ and $\mc{Q}\hat{\rho}(0) = 0$. The evolution of $\mc{P}\hat{\rho}(t)$ under second-order perturbation theory follows\cite{kim2024general1,yang_influence_2002,mulvihill_road_2021}
\begin{equation}\label{eq:2nd_order}
        \frac{d}{dt} \big[ \mc{P} \hat{\rho}(t) \big] \approx -\frac{1}{\hbar^2} \int_0^t \mc{P} \mc{L}_1 \exp \bigg[ -\frac{i(t - \tau)}{\hbar} \mc{L}_0 \bigg] \mc{L}_1 \mc{P} \hat{\rho} (\tau) \: d\tau,
\end{equation}
to which we make a substitution $t - \tau = t'$ and apply Markov approximation by replacing $\hat{\rho}(t - t')$ with $\hat{\rho}(t)$ and extending the upper limit of the integration to infinity. After calculating
$\tr_\ti{b} [ \bra{\alpha} \frac{d}{dt} \{ \mc{P} \hat{\rho}(t) \} \ket{\alpha} ]$ from the resulting expression, we obtain a time-local equation-of-motion for the exciton populations
\begin{equation}\label{eq:rate_prim}
\begin{split}
        \dot{P}_\alpha(t) = -\frac{1}{\hbar^2} \tr_\ti{b} \bigg[ \int_0^\infty \bra{\alpha} \mc{P} \mc{L}_1 \exp( -it' \mc{L}_0 / \hbar ) \\\mc{L}_1 \mc{P} \hat{\rho}(t) \ket{\alpha} \: dt' \bigg].
\end{split}
\end{equation}
Expanding the exponential in \eq{eq:rate_prim} leads to a first-order rate equation
\begin{equation}\label{eq:rateeqn_pop}
    \dot{P}_\alpha(t) = \sum_{\beta \neq \alpha} [-K_{\beta \alpha} P_\alpha(t) + K_{\alpha \beta} P_\beta(t)],
\end{equation}
where the rate constants are expressed as
\begin{equation}\label{eq:rateconst_int}
    K_{\beta \alpha} = \frac{2}{\hbar^2} \ti{Re} \int_0^\infty \exp \left( \frac{-it'(E_\beta - E_\alpha)}{\hbar} \right) S_{\beta \alpha} (t') \: dt',
\end{equation}
\begin{equation}\label{eq:trace_bath_full}
    S_{\beta \alpha} (t') = \tr_\ti{b} \big[ \hat{U}_\alpha^\dagger(t') \hat{V}_{\alpha\beta} \hat{U}_\beta (t') \hat{V}_{\beta\alpha} \hat{R}_\alpha \big],
\end{equation}
with the time-dependent unitary operators $\{\hat{U}_\alpha (t')\}$ defined by
\begin{equation}\label{eq:unitary}
    \hat{U}_\alpha(t') = \exp \bigg( - \frac{i t' \hat{V}_{\alpha \alpha}}{\hbar} \bigg).
\end{equation}
The condition that the integral in \eq{eq:rateconst_int} is well-defined
\begin{equation}\label{eq:S_limit}
    \lim_{t' \rightarrow \infty} S_{\beta \alpha}(t') = 0,
\end{equation}
will play a crucial role in the proof of energy conservation in \stn{section:econserv}. Integrating \eq{eq:rateconst_int} requires us to accurately determine $S_{\beta \alpha}(t')$ up to an arbitrary time point. For this, we factorize the unitary operator [\eq{eq:unitary}] and equilibrium bath density [\eq{eq:bathdensity}] into contributions from individual bath components,
\begin{equation}
    \hat{U}_\alpha (t') = \prod_j \hat{u}_{\alpha}^j(t'), \quad \hat{u}_{\alpha}^j(t') = \exp \bigg( - \frac{it' \hat{v}_{\alpha\alpha}^j}{\hbar} \bigg),
\end{equation}
and
\begin{equation}
    \hat{R}_\alpha = \prod_j \hat{r}_{\alpha}^j, \quad \hat{r}_{\alpha}^j = \frac{\exp(-\beta \hat{v}_{\alpha\alpha}^j)}{\tr_j [\exp(-\beta \hat{v}_{\alpha\alpha}^j)]}.
\end{equation}
Here, $\tr_j$ indicates the trace over the subspace spanned by the $j$th bath component. For succinctness, the dependence on $t'$ of the scalars and operators will be omitted hereafter, unless required for clarity. 

We now define the abbreviation for the traces
\begin{subequations}\label{eq:traces}
    \begin{equation}\label{eq:traces-a}
    \tr0_{\beta \alpha}^j \equiv \tr_j \big[ (\hat{u}_{\alpha}^j)^\dagger \hat{u}_{\beta}^j \hat{r}_{\alpha}^j \big],
    \end{equation}
    \begin{equation}\label{eq:traces-b}
    \tr1_{\beta \alpha}^j \equiv \tr_j \big[ (\hat{u}_{\alpha}^j)^\dagger  \hat{v}_{\alpha\beta}^j \hat{u}_{\beta}^j \hat{r}_{\alpha}^j \big],
    \end{equation}
    \begin{equation}\label{eq:traces-c}
    \tr2_{\beta \alpha}^j \equiv \tr_j \big[ (\hat{u}_{\alpha}^j)^\dagger  \hat{u}_{\beta}^j \hat{v}_{\beta\alpha}^j \hat{r}_{\alpha}^j \big],
    \end{equation}
    \begin{equation}\label{eq:traces-d}
    \tr3_{\beta \alpha}^j \equiv \tr_j \big[ (\hat{u}_{\alpha}^j)^\dagger \hat{v}_{\alpha\beta}^j \hat{u}_{\beta}^j \hat{v}_{\beta\alpha}^j \hat{r}_{\alpha}^j \big].
    \end{equation}
\end{subequations}
By adopting this notation, the trace in \eq{eq:trace_bath_full} can be expressed as
\begin{equation}\label{eq:trace_const}
\begin{split}
        S_{\beta \alpha}(t') =& \sum_j \bigg[ \tr3_{\beta \alpha}^j \prod_{k \neq j} \tr0_{\beta \alpha}^k +\\ &\sum_{k \neq j} \bigg( \tr1_{\beta \alpha}^j \tr2_{\beta \alpha}^k \prod_{l \neq (j, k)} \tr0_{\beta \alpha}^l \bigg) \bigg].
\end{split}
\end{equation}
Because we did not adopt any specific model of the bath up to this point, \eq{eq:trace_const} is valid
for arbitrary bath and system-bath interaction, as long as they can be decomposed into the form of \eq{eq:Hbath_int_sum}. Importantly, \eq{eq:trace_const} disentangles the trace for the full bath subspace [\eq{eq:trace_bath_full}] into the traces for individual bath components [\eq{eq:traces}].
To numerically evaluate \eq{eq:traces} for all bath components we can convert \eq{eq:trace_const} to a more practical expression by defining
\begin{equation}\label{eq:trace_sum_jum}
    \begin{gathered}
    W_{\beta \alpha} \equiv \sum_j \frac{\tr1_{\beta \alpha}^j}{\tr0_{\beta \alpha}^j}, \quad X_{\beta \alpha} \equiv \sum_j \frac{\tr2_{\beta \alpha}^j}{\tr0_{\beta \alpha}^j},\\ \ Y_{\beta \alpha} \equiv \sum_j \frac{\tr3_{\beta \alpha}^j}{\tr0_{\beta \alpha}^j},\quad 
    \Pi_{\beta \alpha} \equiv \prod_j \tr0_{\beta \alpha}^j, \\ Z_{\beta \alpha} \equiv \sum_j \frac{\tr1_{\beta \alpha}^j \tr2_{\beta \alpha}^j}{(\tr0_{\beta \alpha}^j)^2},
    \end{gathered}
\end{equation}
such that
\begin{equation}\label{eq:trace_const_jum}
    S_{\beta \alpha}(t') = (W_{\beta \alpha} X_{\beta \alpha} + Y_{\beta \alpha} - Z_{\beta \alpha}) \Pi_{\beta \alpha}.
\end{equation}
\Eqs{eq:trace_sum_jum}{eq:trace_const_jum} evaluates $S_{\beta \alpha}(t')$ at a computational cost proportional to $\mc{O}(n)$, instead of the naive implementation of \eq{eq:trace_const} which scales as $\mc{O}(n^2)$ due to the existence of the double summation.

As we will demonstrate in \stn{subsubsection:hobath_pop}, for relatively simple bath models such as harmonic oscillators with linear system-bath coupling, it is even possible to condense \eq{eq:trace_const_jum} into a single analytical expression. However, there may also be situations where this simplification is not feasible. In such cases, we can utilize an incremental formula
\begin{equation}\label{eq:recur_pop}
\begin{split}
        S_{\beta \alpha}(t') =& \big[ \tr3_{\beta \alpha}^j + (\tr2_{\beta \alpha}^j) W_{\beta \alpha}^{j-} + (\tr1_{\beta \alpha}^j) X_{\beta \alpha}^{j-} \big] \Pi_{\beta \alpha}^{j-} \\&+ (\tr0_{\beta \alpha}^j) S_{\beta \alpha}^{j-}(t'),
\end{split}
\end{equation}
where the quantities with the subscript $j-$ are similarly defined as in \eq{eq:trace_const_jum} but they exclude the contribution from the $j$th bath component,
\begin{equation}\label{eq:trace_sum_jum_j}
    \begin{gathered}
    W_{\beta \alpha}^{j-} \equiv \sum_{k\neq j} \frac{\tr1_{\beta \alpha}^k}{\tr0_{\beta \alpha}^k}, \quad X_{\beta \alpha}^{j-} \equiv \sum_{k\neq j} \frac{\tr2_{\beta \alpha}^k}{\tr0_{\beta \alpha}^k}, \\ Y_{\beta \alpha}^{j-} \equiv \sum_{k\neq j} \frac{\tr3_{\beta \alpha}^k}{\tr0_{\beta \alpha}^k},\quad 
    \Pi_{\beta \alpha}^{j-} \equiv \prod_{k\neq j} \tr0_{\beta \alpha}^k,\\
    Z_{\beta \alpha}^{j-} \equiv \sum_{k\neq j} \frac{\tr1_{\beta \alpha}^k \tr2_{\beta \alpha}^k}{(\tr0_{\beta \alpha}^k)^2}.
    \end{gathered}
\end{equation}
By using \eq{eq:recur_pop} we can efficiently calculate $S_{\beta \alpha}(t')$ by successively incorporating the effect of problematic components to the analytical expression already representing most of the bath, rather than immediately retreating to the direct application of \eq{eq:trace_sum_jum} and \eq{eq:trace_const_jum}. 

\subsection{Dissipation}
\label{subsection:dissipation}
To quantify the dissipation into individual bath components [\eq{eq:Hbath_int_sum}], we need to evaluate the rate of dissipation for the $j$th bath component as\cite{kim2024general1}
\begin{equation}\label{eq:equivalence}
    \dot{E}_j (t) = \tr \bigg[ \hat{h}_j \frac{d}{dt} \big[ \mc{P}_{j-} \hat{\rho}(t) \big] \bigg].
\end{equation}
\Eq{eq:equivalence} features a new projection operator $\mc{P}_{j-}$ which satisfy $\mc{P} = \hat{p}_j \mc{P}_{j-}$, where
\begin{equation}\label{eq:proj_j}
    \hat{p}_j \hat{\rho} = \sum_\alpha \bigg( \tr_j [\bra{\alpha} \hat{\rho} \ket{\alpha} ] \ket{\alpha} \bra{\alpha} \otimes \hat{r}_{\alpha}^j \bigg),
\end{equation}
\begin{equation}\label{eq:proj_excl}
    \mc{P}_{j-} \hat{\rho} = \sum_\alpha \bigg( \tr_{\ti{b}}^{j-} [\bra{\alpha} \hat{\rho} \ket{\alpha} ] \ket{\alpha} \bra{\alpha} \otimes \hat{R}_{\alpha}^{j-} \bigg).
\end{equation}
In the above, $\tr_{\ti{b}}^{j-}$ denotes the trace over the subspace of all bath components except the $j$th component, and $\hat{R}_{\alpha}^{j-}$ is the equilibrium bath density in this subspace
\begin{equation}
    \hat{R}_{\alpha}^{j-} = \prod_{k \neq j} \hat{r}_{\alpha}^{k}.
\end{equation}
Employing $\mc{P}_{j-}$ [\eq{eq:proj_excl}] in \eq{eq:equivalence} removes the projection for the $j$th bath component, which is crucial for quantifying the dissipation by this component after an infinitesimal amount of time.\cite{kim2024general1} After calculating the dissipation, the system density returns to the fully projected form $\hat{\mc{P}}\hat{\rho}$ by applying the remaining part of the projection operator $\hat{p}_j$ [\eq{eq:proj_j}], achieving consistency with the population dynamics governed by \eq{eq:rateeqn_pop}.

We aim to develop a practical method for evaluating \eq{eq:equivalence}. We start by observing that the time-evolution of $\mc{P}_{j-} \hat{\rho}(t)$ under the second-order perturbation theory follows the equation of motion similar to \eq{eq:2nd_order} except $\mc{P}$ is replaced by $\mc{P}_{j-}$,\cite{kim2024general1}
\begin{equation}\label{eq:2nd_order_excl}
    \frac{d}{dt} \big[ \mc{P}_{j-} \hat{\rho}(t) \big] = -\frac{1}{\hbar^2} \int_0^t \mc{P}_{j-} \mc{L}_1 \exp \bigg[ -\frac{i(t - \tau)}{\hbar} \mc{L}_0 \bigg] \mc{L}_1 \mc{P}_{j-} \hat{\rho} (\tau) \: d\tau.
\end{equation}
Applying the Markov approximation gives
\begin{equation}\label{eq:2nd_order_Markov}
    \frac{d}{dt} \big[ \mc{P}_{j-} \hat{\rho}(t) \big] = -\frac{1}{\hbar^2} \int_0^\infty \mc{P}_{j-} \mc{L}_1 \exp( -it'\mc{L}_0 / \hbar) \mc{L}_1 \mc{P}_{j-} \hat{\rho} (t) \: dt'.
\end{equation}
Because we are focusing on the evolution of $\mc{P} \hat{\rho}(t)$, it is valid to assume that $\hat{\rho}(t) = \mc{P} \hat{\rho}(t)$ is satisfied at every instance. Under this circumstance, the integrand of \eq{eq:2nd_order_Markov} can be expanded as
\begin{equation}
    \begin{split}
    \mc{P}_{j-} &\mc{L}_1 \exp( -it'\mc{L}_0 / \hbar) \mc{L}_1 \mc{P}_{j-} = \sum_\alpha \sum_{\beta \neq \alpha} \bigg[  
    \exp \left( -\frac{it' (E_\beta - E_\alpha)}{\hbar} \right)\\
    &\ket{\alpha} \bra{\alpha} \otimes \hat{R}_{\alpha}^{j-} \otimes \bigg( P_\alpha(t) \tr_{\ti{b}}^{j-} \big[ \hat{V}_{\alpha\beta} \hat{U}_\beta \hat{V}_{\beta\alpha} \hat{R}_\alpha \hat{U}_\alpha^\dagger \big] \\&- P_\beta(t) \tr_{\ti{b}}^{j-} \big[ \hat{V}_{\alpha\beta} \hat{U}_\beta \hat{R}_\beta \hat{V}_{\beta\alpha} \hat{U}_\alpha^\dagger \big] \bigg) \bigg] + \ti{H.c.},
    \end{split}
\end{equation}
where the traces on the right-hand side are now operators related to the $j$th component, rather than scalars as in \eq{eq:rateconst_int}.
We now switch  $\alpha$ and $\beta$ for the two terms involving $P_\beta(t)$ on the right-hand side of \eq{eq:2nd_order_Markov}, which is justified by the fact that the summation is over all ordered pairs of $\alpha$ and $\beta$. The resulting expression can then be used with 
\eq{eq:Hbathcomp_exci} to evaluate the right-hand side of \eq{eq:equivalence}, leading to a first-order rate equation for the dissipation  
\begin{equation}\label{eq:rateeqn_diss}
    \dot{E}_j(t) = \sum_\alpha \sum_{\beta \neq \alpha} \mathcal{K}_{\beta \alpha}^j P_\alpha(t),
\end{equation}
with the rate constants given by
\begin{equation}\label{eq:rateconst_diss_int}
    \begin{split}
    \mathcal{K}_{\beta \alpha}^j &= \frac{2}{\hbar^2} \: \ti{Re} \int_0^\infty \exp \bigg( -\frac{i t' (E_\beta - E_\alpha)}{\hbar} \bigg) \mathcal{S}_{\beta \alpha}^j(t') \: dt',
    \end{split}
\end{equation}
\begin{equation}\label{eq:trace_bath_diss_full}
\begin{split}
        \mathcal{S}_{\beta \alpha}^j(t') &= \tr_\ti{b} \big[ \hat{v}_{\beta\beta}^{j} \hat{U}_\beta \hat{V}_{\beta\alpha} \hat{R}_\alpha \hat{U}_\alpha^\dagger \hat{V}_{\alpha\beta} \big] \\ &- \tr_\ti{b} \big[ \hat{v}_{\alpha\alpha}^{j} \hat{V}_{\alpha\beta} \hat{U}_\beta \hat{V}_{\beta\alpha} \hat{R}_\alpha \hat{U}_\alpha^\dagger \big].
\end{split}
\end{equation}
As for the population transfer rate constants $\{K_{BA}\}$ [\eq{eq:rateconst_int}], explicit evaluation of \eq{eq:rateconst_diss_int} requires disassembling $\mathcal{S}_{\beta \alpha}^j(t')$ [\eq{eq:trace_bath_diss_full}] into contributions arising from individual bath components. For this purpose, we extend the shorthand notation introduced in \eq{eq:traces} by additionally defining 
\begin{subequations}\label{eq:traces2}
\begin{equation}
\begin{split}
        \tr4_{\beta \alpha}^j &\equiv \tr_j \big[ (\hat{u}_\alpha^j)^\dagger (\hat{v}_{\beta\beta}^j - \hat{v}_{\alpha\alpha}^j) \hat{u}_\beta^j \hat{r}_\alpha^j \big] \\&= i\hbar \frac{d{\tr}0_{\beta\alpha}^j}{dt'},
\end{split}
\end{equation}
\begin{equation}
\begin{split}
        \tr5_{\beta \alpha}^j &\equiv \tr_j \big[ (\hat{u}_\alpha^j)^\dagger (\hat{v}_{\alpha\beta}^j \hat{v}_{\beta\beta}^j - \hat{v}_{\alpha\alpha}^j \hat{v}_{\alpha\beta}^j) \hat{u}_\beta^j \hat{r}_\alpha^j \big] \\&= i\hbar \frac{d{\tr}1_{\beta\alpha}^j}{dt'},
\end{split}
\end{equation}
\begin{equation}
\begin{split}
    \tr6_{\beta \alpha}^j &\equiv \tr \big[ (\hat{u}_\alpha^j)^\dagger (\hat{v}_{\beta\beta}^j - \hat{v}_{\alpha\alpha}^j) \hat{u}_\beta^j \hat{v}_{\beta\alpha}^j \hat{r}_\alpha^j \big] \\&= i\hbar \frac{d{\tr}2_{\beta\alpha}^j}{dt'},
\end{split}
\end{equation}
\begin{equation}
\begin{split}
    \tr7_{\beta \alpha}^j &\equiv \tr_j \big[ (\hat{u}_{\alpha}^j)^\dagger (\hat{v}_{\alpha\beta}^j \hat{v}_{\beta\beta}^j - \hat{v}_{\alpha\alpha}^j \hat{v}_{\alpha\beta}^j) \hat{u}_\beta^j \hat{v}_{\beta\alpha}^j \hat{r}_{\alpha}^j \big] \\&= i\hbar \frac{d{\tr}3_{\beta \alpha}^j}{dt'}.
\end{split}
\end{equation}
\end{subequations}
and express $\mathcal{S}_{\beta \alpha}^j(t')$ in terms of the traces for individual bath components [\eqs{eq:traces}{eq:traces2}]. The resulting expression can be simplified using the abbreviated notation in \eq{eq:trace_sum_jum},
\begin{equation}\label{eq:recur_diss}
    \begin{split}
    \mathcal{S}_{\beta \alpha}^j(t') &= \big[  (\tr6_{\beta \alpha}^j) W_{\beta \alpha}^{j-} + (\tr5_{\beta \alpha}^j) X_{\beta \alpha}^{j-} \big] \Pi_{\beta \alpha}^{j-} \\&+ \tr7_{\beta \alpha}^j +(\tr4_{\beta \alpha}^j) S_{\beta \alpha}^{j-}(t').
    \end{split}
\end{equation}

\subsection{Proof of thermodynamic principles}\label{sec:thermo}
\subsubsection{Energy conservation}\label{section:econserv}
To prove energy conservation, we need to show that the rate of energy loss from the system is equal to the rate of energy gain by the bath,
\begin{equation}\label{eq:econserv_prim}
    \frac{d}{dt} \tr \big[ \hat{H}_\ti{sub} \mc{P}\hat{\rho}(t) \big] + \sum_j \dot{E}_j(t) \overset{?}{=} 0,
\end{equation}
within our scope which focuses on $\mc{P}\hat{\rho}(t)$. 

We eliminate the time-derivatives in \eq{eq:econserv_prim} by invoking Eq.~(\ref{eq:Hsub_exci}), (\ref{eq:proj_op}), (\ref{eq:rateeqn_pop}), and (\ref{eq:rateeqn_diss}), and then rearrange the resulting expression to get
\begin{equation}\label{eq:econserv}
    \sum_\alpha \sum_{\beta \neq \alpha} \bigg( (E_\beta - E_\alpha) K_{\beta \alpha} + \sum_j \mathcal{K}_{\beta \alpha}^j \bigg) P_\alpha(t) \overset{?}{=} 0.
\end{equation}
The requirement for \eq{eq:econserv} to be satisfied for arbitrary set of populations $\{ P_\alpha(t) \}$ is
\begin{equation}\label{eq:econserv_lowlv}
    (E_\beta - E_\alpha) K_{\beta \alpha} + \sum_j \mathcal{K}_{\beta \alpha}^j \overset{?}{=} 0,
\end{equation}
for any pairs of $\alpha$ and $\beta$. Replacing the population transfer and dissipation rate constants with their explicit expressions [Eqs.~(\ref{eq:rateconst_int}), (\ref{eq:trace_bath_full}), (\ref{eq:rateconst_diss_int}), and (\ref{eq:trace_bath_diss_full})] gives
\begin{equation}\label{eq:econserv_interm}
    \begin{split}
    (E_\beta &- E_\alpha) K_{\beta \alpha} + \sum_j \mathcal{K}_{\beta \alpha}^j = 
    \frac{2}{\hbar^2} \: \ti{Re} \int_0^\infty  \\
    &\times \bigg( \tr_\ti{b} \big[ (E_\beta + \hat{V}_{\beta \beta}) \hat{U}_\beta \hat{V}_{\beta \alpha} \hat{R}_\alpha \hat{U}_\alpha^\dagger \hat{V}_{\alpha\beta} \big] \\
    &- \tr_\ti{b} \big[ (E_\alpha + \hat{V}_{\alpha\alpha}) \hat{V}_{\alpha\beta} \hat{U}_\beta \hat{V}_{\beta\alpha} \hat{R}_\alpha \hat{U}_\alpha^\dagger \big] \bigg) \\&\exp \bigg( -\frac{it' (E_\beta - E_\alpha)}{\hbar} \bigg)\: dt',
    \end{split}
\end{equation}
where we used \eq{eq:bath_exci_component} to condense the sum of the operators for individual bath components. Then, we invoke \eqs{eq:trace_bath_full}{eq:unitary} to express the integrand on the right-hand side of \eq{eq:econserv_interm} as a time-derivative,
\begin{equation}\label{eq:econserv_interm2}
     \begin{split}
    (E_\beta - E_\alpha) K_{\beta \alpha} &+ \sum_j \mathcal{K}_{\beta \alpha}^j = 
    \frac{2}{\hbar^2} \: \ti{Re} \int_0^\infty i\hbar \frac{d}{dt'} \\& \exp \bigg( -\frac{it' (E_\beta - E_\alpha)}{\hbar} \bigg) S_{\beta\alpha}(t')  \: dt'.
    \end{split}
\end{equation}
We can now carry out the integration and simplify the result with $\hat{U}_\alpha(0) = 1$ and \eq{eq:S_limit} to obtain
\begin{equation}
    \int_0^\infty \bigg( i\hbar \frac{d}{dt'} \tr_\ti{b} \big[ \hat{U}_\alpha^\dagger \hat{V}_{\alpha\beta} \hat{U}_\beta \hat{V}_{\beta\alpha} \hat{R}_\alpha \big] \bigg) \: dt' = i \hbar \: \tr_\ti{b} \big[ \hat{V}_{\alpha\beta} \hat{V}_{\beta\alpha} \hat{R}_\alpha \big],
\end{equation}
whose value is purely imaginary as $\tr_\ti{b} [ \hat{V}_{\alpha\beta} \hat{V}_{\beta\alpha} \hat{R}_\alpha ] = \tr_\ti{b} [ (\hat{V}_{\alpha\beta} \hat{V}_{\beta\alpha} \hat{R}_\alpha)^\dagger]$ is real. As a result, the right-hand side of \eq{eq:econserv_interm2} vanishes and assures the validity of \eq{eq:econserv} and, in turn, \eq{eq:econserv_prim}. Therefore, we can conclude that the dissipation calculated by Eqs.~(\ref{eq:rateeqn_diss})--(\ref{eq:trace_bath_diss_full}) satisfies the energy conservation and achieves consistency with the population dynamics.

\subsubsection{Detailed balance}\label{subsubsection:detbal}

For the dynamics of population and dissipation governed by \eqs{eq:rateeqn_pop}{eq:rateeqn_diss}, the detailed balance condition is represented as
\begin{equation}\label{eq:diss_detbal}
    - \frac{\mc{K}_{\alpha\beta}^j}{\mc{K}_{\beta\alpha}^j} = \frac{K_{\alpha\beta}}{K_{\beta\alpha}} = \frac{P_\alpha(\infty)}{P_\beta(\infty)},
\end{equation}
which makes the net dissipation by any bath component vanish at the steady state. To prove \eq{eq:diss_detbal}, we start by applying the Wick rotation $t' \to t' -i\hbar\beta$ to $S_{\beta\alpha}(t')$ [\eq{eq:trace_bath_full}],
\begin{equation}
\begin{split}
        S_{\beta\alpha} (t' - i \hbar \beta) = \tr_\ti{b} \big[& \exp(\beta \hat{V}_{\alpha \alpha}) \hat{U}_\alpha^\dagger \hat{V}_{\alpha \beta} \hat{U}_\beta \\&\exp(- \beta \hat{V}_{\beta \beta}) \hat{V}_{\beta \alpha} \hat{R}_\alpha \big],
\end{split}
\end{equation}
and rearrange the right-hand side to get
\begin{equation}\label{eq:pop_conj}
    S_{\beta\alpha}(t' - i \hbar \beta) = \frac{\tr_\ti{b} [\exp(- \beta \hat{V}_{\beta \beta})]}{\tr_\ti{b} [\exp(- \beta \hat{V}_{\alpha \alpha})]} \big[ S_{\alpha\beta}(t') \big]^*,
\end{equation}
which can be readily validated by using the cyclic invariance of the trace and the definition of the thermal bath density [\eq{eq:bathdensity}]. If we define the Fourier transform of $S_{\beta \alpha}(t')$ as $\tilde{S}_{\beta \alpha} (\omega)$, it can be shown with \eq{eq:pop_conj} that the population transfer rates in the opposite directions can be expressed as
\begin{subequations}\label{eq:epsilon_eq04}
    \begin{equation}
    \begin{split}
    K_{\beta\alpha} = \frac{2}{\hbar^2} \: \tilde{S}_{\beta\alpha} \bigg( \frac{E_\beta - E_\alpha}{\hbar} \bigg),
    \end{split}
    \end{equation}
    \begin{equation}
    \begin{split}
    K_{\alpha\beta} = \frac{2}{\hbar^2} \frac{\tr_\ti{b} [\exp\{-\beta (E_\alpha + \hat{V}_{\alpha\alpha}) \}]}{\tr_\ti{b} [\exp\{-\beta (E_\beta + \hat{V}_{\beta\beta}) \}]} \tilde{S}_{\beta\alpha} \bigg( \frac{E_\beta - E_\alpha}{\hbar} \bigg).
    \end{split}
    \end{equation}
\end{subequations}
Hence, the ratio between the two rate constants becomes
\begin{equation}\label{eq:detbal_op}
    \frac{K_{\alpha\beta}}{K_{\beta\alpha}} = \frac{\tr_\ti{b} [\exp\{-\beta (E_\alpha + \hat{V}_{\alpha\alpha}) \}]}{\tr_\ti{b} [\exp\{-\beta (E_\beta + \hat{V}_{\beta\beta}) \}]}.
\end{equation}

We then move onto the dissipation and apply a similar procedure to $\mc{S}_{\beta\alpha}^j(t')$ [\eq{eq:trace_bath_diss_full}] to deduce
\begin{equation}\label{eq:diss_conj}
    \mc{S}_{\beta\alpha}(t' - i \hbar \beta) = - \frac{\tr_\ti{b} [\exp(- \beta \hat{V}_{\beta \beta})]}{\tr_\ti{b} [\exp(- \beta \hat{V}_{\alpha \alpha})]} \big[ \mc{S}_{\alpha\beta}(t') \big]^*,
\end{equation}
which leads to
\begin{equation}\label{eq:detbal_diss}
    \frac{\mc{K}_{\alpha\beta}^j}{\mc{K}_{\beta\alpha}^j} = - \frac{\tr_\ti{b} [\exp\{-\beta (E_\alpha + \hat{V}_{\alpha\alpha}) \}]}{\tr_\ti{b} [\exp\{-\beta (E_\beta + \hat{V}_{\beta\beta}) \}]}.
\end{equation}
\Eq{eq:diss_detbal} is now instantly validated by combining \eqs{eq:detbal_op}{eq:detbal_diss}.

\subsection{Application to linearly  coupled harmonic oscillator Bath}\label{subsection:hobath}

As a concrete example, we apply the framework developed in \stn{subsection:population} and \stn{subsection:dissipation}
to analyze the dissipation by a bath of quantum harmonic oscillators. In this case, the bath Hamiltonian takes the form 
\begin{equation}\label{eq:Hbath_lho}
    \hat{H}_\ti{B} = \sum_j \bigg( \frac{\hat{p}_j^2}{2} + \frac{\omega_j^2 \hat{x}_j^2}{2} \bigg),
\end{equation}
where $\hat{p}_j$ and $\hat{x}_j$ are the mass-weighted momentum and position operators for the $j$th bath mode, and $\omega_j$ is the characteristic frequency. We assume that the coupling between the system and individual bath modes linearly depends on the positional coordinates, such that
\begin{equation}\label{eq:Hint_lho}
    \hat{H}_\ti{SB} = - \sum_{A}\sum_{B} \bigg( \ket{A}\bra{B} \otimes \sum_j (\omega_j^2 d_{AB}^j \hat{x}_j + \gamma_{AB}^j) \bigg),
\end{equation}
where $d_{AB}^j$ determines the strength of the system-bath interaction and $\gamma_{AB}^j$ accounts for the possible energy shift that arises from the freedom of setting the boundary between the system and the bath [\eqs{eq:Hfull}{eq:Hsub_dia}]. The profile of the system-bath coupling in the frequency domain is contained in the spectral densities
\begin{equation} \label{eq:spd-site}
    J_{AB, CD}(\omega) = \sum_j \frac{\omega_j^3 d_{AB}^j d_{CD}^j}{2} \delta(\omega - \omega_j).
\end{equation}
which can take into account both independent ($A = C \text{ and } B = D$) and correlated ($A \neq C \text{ or } B \neq D$) quantum fluctuations induced by the system-bath interaction.

By converting \eqs{eq:Hbath_lho}{eq:Hint_lho} to the exciton basis according to \eq{eq:exc-basis}, we can specify the form of the bath-related operators in \eq{eq:v_exci} as
\begin{equation}\label{eq:v_exci_ho}
    \hat{v}_{\alpha\beta}^j = \bigg( \frac{\hat{p}_j^2}{2} + \frac{\omega_j^2 \hat{x}_j^2}{2} \bigg) \delta_{\alpha\beta} - \omega_j^2 d_{\alpha\beta}^j \hat{x}_j + \gamma_{\alpha\beta}^j,
\end{equation}
where $\delta_{\alpha \beta}$ is the Kronecker's delta and
\begin{subequations}\label{eq:d_exci}
\begin{equation}
    d_{\alpha\beta}^j = \sum_{A}\sum_{B} c_{\alpha A}^* c_{\beta B} d_{AB}^j,
\end{equation}
\begin{equation}
    \gamma_{\alpha\beta}^j = \sum_{A}\sum_{B} c_{\alpha A}^* c_{\beta B} \gamma_{AB}^j,
\end{equation}
\end{subequations}
are the coupling strengths and energy shifts in the exciton basis.

\subsubsection{Population transfer}\label{subsubsection:hobath_pop}
Based on \eq{eq:v_exci_ho}, the rate constants for population transfer rate [\eq{eq:trace_const}] and dissipation [\eq{eq:rateconst_diss_int}] can be computed by following the procedure illustrated in \stn{subsection:population} and \stn{subsection:population}, respectively. 

To simplify the expressions that will appear in the  derivations, we take an exciton state $\ket{\alpha}$ as a reference and redefine the positional coordinate according to $\hat{y}_j = \hat{x}_j - d_{\alpha \alpha}^j$ so that the origin $\hat{y}_j = 0$ coincides with the minimum of the PES $\hat{v}_{\alpha \alpha}^j$. In this new coordinate, \eq{eq:v_exci_ho} transforms into four different forms depending on which part of the system the bath-related operators couples to,
\begin{subequations}\label{eq:vop_exci}
\begin{equation}\label{eq:vop_exci-a}
    \hat{v}_{\alpha\alpha}^j = \frac{\hat{p}_j^2}{2} + \frac{\omega_j^2 \hat{y}_j^2}{2} - \lambda_{\alpha \alpha, \alpha \alpha}^j + \gamma_{\alpha \alpha}^j,
\end{equation}
\begin{equation}\label{eq:vop_exci-b}
    \hat{v}_{\beta\beta}^j = \frac{\hat{p}_j^2}{2} + \frac{\omega_j^2}{2} [\hat{y}_j - (d_{\beta\beta}^j - d_{\alpha\alpha}^j)]^2 - \lambda_{\beta\beta, \beta\beta}^j + \gamma_{\beta\beta}^j,
\end{equation}
\begin{equation}\label{eq:vop_exci-c}
    \hat{v}_{\alpha\beta}^j = -\omega_j^2 d_{\alpha\beta}^j \hat{y}_j - 2 \lambda_{\alpha\beta, \alpha\alpha}^j + \gamma_{\alpha\beta}^j,
\end{equation}
\begin{equation}\label{eq:vop_exci-d}
    \hat{v}_{\beta\alpha}^j = -\omega_j^2 d_{\beta\alpha}^j \hat{y}_j - 2 \lambda_{\beta\alpha, \alpha\alpha}^j + \gamma_{\beta\alpha}^j,
\end{equation}
\end{subequations}
where $\alpha \neq \beta$ and $\lambda_{\mu\nu, \xi\chi}^j = \frac{\omega_j^2 d_{\mu\nu}^j d_{\xi\chi}^j}{2}$.

To obtain the rate constants for exciton population transfer [\eq{eq:rateconst_int}] we need to compute $S_{\beta \alpha} (t')$ [\eq{eq:trace_const}], which requires evaluating the traces in \eq{eq:traces} using the bath-related operators defined in \eq{eq:vop_exci}. To evaluate these traces, we begin with $\tr0_{\beta\alpha}^j$ [Eq.~(\ref{eq:traces}a)] whose analytical expression, 
\begin{equation}\label{eq:tr0_ana}
    \begin{split}
    \tr0_{\beta\alpha}^j &= \exp \bigg( -\frac{it'(\Delta_{\beta\alpha}^j + G_{\beta\alpha}^j)}{\hbar} - \frac{G_{\beta\alpha}^j}{\hbar} f(\omega_j, t') \bigg),
    \end{split}
\end{equation}
was obtained using the generalized cumulant expansion technique\cite{sung_four_2001,mukamel_nonimpact_1983} or the small polaron transformation\cite{jang_multistep_2012,jang_nonequilibrium_2002}. Here,  $G_{\beta\alpha}^j$, $\Delta_{\beta\alpha}^j$, and $f(\omega, t')$ are defined as
\begin{equation}
    G_{\beta\alpha}^j = \lambda_{\alpha\alpha, \alpha\alpha}^j - 2 \lambda_{\alpha\alpha, \beta\beta}^j + \lambda_{\beta\beta, \beta\beta}^j,
\end{equation}
\begin{equation}
    \Delta_{\beta\alpha}^j = \lambda_{\alpha\alpha, \alpha\alpha}^j - \lambda_{\beta\beta, \beta\beta}^j - \gamma_{\alpha\alpha}^j + \gamma_{\beta\beta}^j,
\end{equation}
\begin{equation}
    f(\omega, t') = \coth \bigg( \frac{\beta \hbar \omega}{2} \bigg) \frac{1 - \cos (\omega t')}{\omega} + i \frac{\sin(\omega t') - \omega t'}{\omega}.
\end{equation}
As shown in \app{sec:app_traces} the analytical expression for the rest of the traces in \eq{eq:traces} can be obtained as,
\begin{subequations}\label{eq:tr123_ana}
    \begin{equation}\label{eq:tr123_ana-a}
    \begin{split}
        \tr1_{\beta\alpha}^j = -[ &i (\lambda_{\alpha\beta,\beta\beta}^j - \lambda_{\alpha\beta,\alpha\alpha}^j) \dot{f}(\omega_j, t) 
    \\&+ 2 \lambda_{\alpha\beta, \alpha\alpha}^j - \gamma_{\alpha\beta}^j ] \: \tr0_{\beta\alpha}^j,
    \end{split}
    \end{equation}
    \begin{equation}\label{eq:tr123_ana-b}
    \begin{split}
        \tr2_{\beta\alpha}^j = -[& i (\lambda_{\beta\alpha,\beta\beta}^j - \lambda_{\beta\alpha,\alpha\alpha}^j) \dot{f}(\omega_j, t)
    \\&+ 2 \lambda_{\beta\alpha, \alpha\alpha}^j - \gamma_{\beta\alpha}^j ] \: \tr0_{\beta\alpha}^j,
    \end{split}
    \end{equation}
    \begin{equation}\label{eq:tr123_ana-c}
    \begin{split}
    \tr3_{\beta\alpha}^j &= [ \{ i(\lambda_{\alpha\beta,\beta\beta}^j - \lambda_{\alpha\beta,\alpha\alpha}^j) \dot{f}(\omega_j, t) 
    + 2 \lambda_{\alpha\beta, \alpha\alpha}^j - \gamma_{\alpha\beta}^j \} \\
    &\times \{ i(\lambda_{\beta\alpha,\beta\beta}^j - \lambda_{\beta\alpha,\alpha\alpha}^j) \dot{f}(\omega_j, t) 
    + 2 \lambda_{\beta\alpha, \alpha\alpha}^j - \gamma_{\beta\alpha}^j \} \\
    &+ \hbar \lambda_{\alpha\beta, \beta\alpha}^j \ddot{f}(\omega_j, t') ] \: \tr0_{\beta\alpha}^j.
    \end{split}
    \end{equation}
\end{subequations}
\Eqs{eq:tr0_ana}{eq:tr123_ana} allow us to construct the building blocks for $S_{\beta\alpha}(t')$ [\eq{eq:trace_const_jum}] as,

\begin{subequations}\label{eq:blocks_S}
    \begin{equation}
    \begin{split}
         W_{\beta \alpha}(t') = &-i\hbar \{\dot{g}_{\alpha \beta,\beta\beta}(t') - \dot{g}_{\alpha \beta,\alpha\alpha}(t')   \} \\&- 2\Lambda_{\alpha \beta,\alpha\alpha} +\Gamma_{\alpha \beta},
    \end{split}
    \end{equation}
    \begin{equation}
    \begin{split}
        X_{\beta \alpha}(t') = &-i\hbar \{\dot{g}_{\beta\alpha ,\beta\beta}(t') - \dot{g}_{ \beta\alpha,\alpha\alpha}(t')   \} \\&- 2\Lambda_{ \beta\alpha,\alpha\alpha} +\Gamma_{\beta\alpha },
    \end{split}
    \end{equation}
    \begin{equation}
    \begin{split}
        Y_{\beta \alpha}(t')-Z_{\beta \alpha}(t')=\hbar^{2}\ddot{g}_{\alpha \beta,\beta\alpha}(t'),
    \end{split}
    \end{equation}
    \begin{equation}
    \begin{split}
        \Pi_{\beta \alpha}(t')&=\exp \bigg( -\frac{it'}{\hbar} (2 \Lambda_{\alpha \alpha, \alpha \alpha}-2 \Lambda_{\alpha \alpha, \beta \beta}-\Gamma_{\alpha\alpha}+\Gamma_{\beta \beta})\\
    &-\dot{g}_{\alpha \alpha,\alpha\alpha}(t') +2 \dot{g}_{\alpha \alpha,\beta\beta}(t')-\dot{g}_{\beta\beta,\beta\beta}(t') \bigg),
    \end{split}
    \end{equation}
\end{subequations}
where we have defined the sum of $\lambda_{\mu\nu, \xi\chi}^{j}$ and $\gamma_{\mu\nu}^{j}$ over all bath components as
\begin{equation}
    \Lambda_{\mu\nu, \xi\chi} = \sum_j \lambda_{\mu\nu, \xi\chi}^j, \quad \Gamma_{\mu\nu} = \sum_j \gamma_{\mu\nu}^j,
\end{equation}
respectively, and the exciton line-broadening function 
\begin{equation}\label{eq:lbf_exci}
\begin{split}
        g_{\mu\nu, \xi\chi} (t') &= \frac{1}{\hbar} \sum_j \big[ \lambda_{\mu\nu, \xi\chi}^j f(\omega_j, t') \big] \\&= \frac{1}{\hbar} \int_{-\infty}^\infty \frac{J_{\mu\nu, \xi\chi}(\omega)}{\omega} f(\omega, t') \; d\omega,
\end{split}
\end{equation}
where we have introduced the spectral density in the excitonic basis 
\begin{equation}\label{eq:spd_exci}
    J_{\mu\nu, \xi\chi}(\omega) = \sum_j \frac{\omega_j^3 d_{\mu\nu}^j d_{\xi\chi}^j}{2} \delta(\omega - \omega_j).
\end{equation}
The rate constants for population transfer can now be evaluated by plugging $S_{\beta\alpha}(t')$ [\eqs{eq:trace_const_jum}{eq:blocks_S}] in \eq{eq:rateeqn_pop} and integrating numerically. 

To further check the validity of the above expressions, we show that they correctly reproduce the already known results from modified Redfield theory\cite{yang_influence_2002,zhang_exciton-migration_1998} (MRT) when applied to a system of interacting chromophore molecules. For each chromophore, we only consider the ground and the first electronic excited states, whose energy 
difference (“site energy”) undergoes fluctuations induced by interactions with the harmonic vibrational modes. We then take the diabatic state $\ket{A}$ to describe the situation in which only the chromophore $A$ is electronically excited, while the rest remain in their ground state. The MRT assumes the Condon approximation\cite{condon_nuclear_1928}, which declares that the electronic couplings between the diabatic states are not affected by the vibrational DOFs. This is equivalent to  setting $d_{AB}^{j}=0$ when $A \neq B$, with which \eq{eq:d_exci} reduces to
\begin{equation}
    d_{\alpha\beta}^{j}=\sum_{A}c_{\alpha A}^{*}c_{\beta A}d_{AA}^{j}.
\end{equation}
The MRT also sets the diagonal elements of the system Hamiltonian [\eq{eq:Hsub_dia}] as the vertical excitation energies at the minimum of the ground state PES, which makes $\gamma_{\mu\nu}^{j}=0$ for all the bath modes and subsequently $\Gamma_{\mu \nu}=0$ for all exciton state pairs $\mu$ and $\nu$. Applying these conditions to $S_{\alpha \beta}(t')$ by using of \eq{eq:blocks_S}, the result is
\begin{equation}\label{eq:S-MRT}
    S_{\beta\alpha} (t') = \mc{N}_{\beta \alpha}(t')\exp \bigg( -\frac{2it'}{\hbar} \Lambda_{\alpha\alpha,\alpha\alpha}-g_{\alpha\alpha,\alpha\alpha}(t')-g_{\beta\beta,\beta\beta}(t')\bigg)
\end{equation}
where we have defined $\mc{N}_{\beta \alpha}(t')$ as
\begin{equation}
    \begin{split}
        \mc{N}_{\beta \alpha}(t')&= \exp \bigg(\frac{2it'}{\hbar}\Lambda_{\alpha\alpha,\beta\beta}+2g_{\alpha\alpha,\beta\beta}(t')  \bigg)\times\\  
        &\bigg ( 
        -[\hbar \{\dot{g}_{\alpha\beta, \beta\beta}(t') - \dot{g}_{\alpha\beta, \alpha\alpha}(t') \} - 2i\Lambda_{\alpha\beta, \alpha\alpha} ] \\
    &\times [\hbar \{\dot{g}_{\beta\alpha, \beta\beta}(t') - \dot{g}_{\beta\alpha, \alpha\alpha}(t') \} - 2i\Lambda_{\beta\alpha, \alpha\alpha} ]\\& + \hbar^2 \ddot{g}_{\alpha\beta, \beta\alpha}(t') \bigg) .
    \end{split}
\end{equation}
Inserting \eq{eq:S-MRT} into \eq{eq:rateconst_int} gives
\begin{equation} \label{eq:pop-rate}
    K_{\beta \alpha} = \frac{2}{\hbar^2} \: \ti{Re} \int_0^\infty \mc{F}_\alpha^*(t') \mc{N}_{\beta\alpha}(t') \mc{A}_\beta(t') \: dt',
\end{equation}
where
\begin{subequations}
    \begin{equation}
        \mc{F}_\alpha(t') = \exp \bigg[ -\frac{it' ( E_{\alpha 0} - \Lambda_{\alpha\alpha, \alpha\alpha} )}{\hbar} - g_{\alpha\alpha, \alpha\alpha}^*(t') \bigg],
    \end{equation}
    \begin{equation}
        \mc{A}_\beta(t') = \exp \bigg[ -\frac{it' ( E_{\beta 0} + \Lambda_{\beta\beta, \beta\beta} )}{\hbar} - g_{\beta\beta, \beta\beta}(t') \bigg],
    \end{equation}
    \begin{align}
    \begin{split}
    \mc{N}_{\beta\alpha}(t') &=  \exp \bigg( \frac{2it'}{\hbar} \Lambda_{\alpha\alpha, \beta\beta} + 2 g_{\alpha\alpha, \beta\beta} (t') \bigg)\times\\&\bigg( - \big[ \hbar \{ \dot{g}_{\alpha\beta, \beta\beta} (t') - \dot{g}_{\alpha\beta, \alpha\alpha} (t') \} - 2 i \Lambda_{\alpha\beta, \alpha\alpha} \big] \\
    &\times \big[ \hbar \{ \dot{g}_{\beta\alpha, \beta\beta} (t') - \dot{g}_{\beta\alpha, \alpha\alpha} (t') \} - 2 i \Lambda_{\beta\alpha, \alpha\alpha} \big] \\&+ \hbar^2 \ddot{g}_{\alpha\beta, \beta\alpha} (t') \bigg) ,
    \end{split}
    \end{align}
\end{subequations}
which are in accord with the expressions for MRT reported in Ref. \cite{yang_influence_2002}. Note that we have defined the zero-phonon exciton energies as 
\begin{equation}
    E_{\mu 0} = E_{\mu}-\Lambda_{\mu\mu,\mu\mu}.
\end{equation}

\subsubsection{Dissipation}\label{subsubsection:hobath_diss}
Our next objective is to calculate the dissipation rate constants \eq{eq:rateconst_diss_int}, for which the most crucial quantity is $\mc{S}_{\beta\alpha}^j(t')$ [\eq{eq:trace_bath_diss_full}]. We first insert \eqs{eq:tr0_ana}{eq:tr123_ana} in \eq{eq:traces2} to derive concrete expressions for the traces that are additionally required to calculate the dissipation,
\begin{subequations}\label{eq:traces-HO}
    \begin{equation}
        \ti{Tr}4_{\beta\alpha}^{j}=[\Delta_{\beta\alpha}^{j}+G_{\beta \alpha}^{j}+iG_{\beta \alpha}^{j}\dot{f}(\omega_{j},t')]\ti{Tr}0_{\beta\alpha}^{j} 
    \end{equation}
    \begin{equation}
    \begin{split}
        \ti{Tr}5_{\beta\alpha}^{j}&=[\Delta_{\beta\alpha}^{j}+G_{\beta \alpha}^{j}+iG_{\beta \alpha}^{j}\dot{f}(\omega_{j},t')]\ti{Tr}1_{\beta\alpha}^{j}\\
        &+\hbar(\lambda_{\alpha\beta,\beta\beta}^{j}-\lambda_{\alpha\beta,\alpha\alpha}^{j})\ddot{f}(\omega_{j},t')\ti{Tr}0_{\beta\alpha}^{j}
    \end{split} 
    \end{equation}
    \begin{equation}
    \begin{split}
        \ti{Tr}6_{\beta\alpha}^{j}&=[\Delta_{\beta\alpha}^{j}+G_{\beta \alpha}^{j}+iG_{\beta \alpha}^{j}\dot{f}(\omega_{j},t')]\ti{Tr}2_{\beta\alpha}^{j}\\
        &+\hbar(\lambda_{\beta\alpha,\beta\beta}^{j}-\lambda_{\beta\alpha,\alpha\alpha}^{j})\ddot{f}(\omega_{j},t')\ti{Tr}0_{\beta\alpha}^{j} 
    \end{split}
    \end{equation}
    \begin{equation}
        \begin{split}
            \ti{Tr}7_{\beta\alpha}^{j}&=[\Delta_{\beta\alpha}^{j}+G_{\beta \alpha}^{j}+iG_{\beta \alpha}^{j}\dot{f}(\omega_{j},t')]\ti{Tr}3_{\beta\alpha}^{j}\\&+\hbar(\lambda_{\alpha\beta,\beta\beta}^{j}-\lambda_{\alpha\beta,\alpha\alpha}^{j})\ddot{f}(\omega_{j},t')\ti{Tr}2_{\beta\alpha}^{j} \\
            &+\hbar(\lambda_{\beta\alpha,\beta\beta}^{j}-\lambda_{\beta\alpha,\alpha\alpha}^{j})\ddot{f}(\omega_{j},t')\ti{Tr}1_{\beta\alpha}^{j} \\
            &+i\hbar^{2}\lambda_{\alpha\beta,\beta\alpha}f^{(3)}(\omega_{j},t')\ti{Tr}0_{\beta\alpha}^{j} 
        \end{split}
    \end{equation}
\end{subequations} 
where the traces in the right-hand sides of the equations are kept in their abbreviated form for compactness. If we substitute the traces in \eq{eq:recur_diss} with the corresponding expressions in \eq{eq:traces-HO}, it can be noticed that some simplifications can be made by utilizing \eq{eq:recur_pop} and 
\begin{subequations}
\begin{equation}    (\ti{Tr}0_{\beta\alpha}^{j}X_{\beta\alpha}^{j-}+\ti{Tr}1_{\beta\alpha}^{j})\Pi_{\beta\alpha}^{j-}=X_{\beta\alpha}\Pi_{\beta\alpha},
\end{equation}
\begin{equation}
(\ti{Tr}0_{\beta\alpha}^{j}W_{\beta\alpha}^{j-}+\ti{Tr}2_{\beta\alpha}^{j})\Pi_{\beta\alpha}^{j-}=W_{\beta\alpha}\Pi_{\beta\alpha},
\end{equation}
\end{subequations}
which can be deduced from \eqs{eq:trace_sum_jum}{eq:trace_sum_jum_j}. As a results, we get
\begin{equation}\label{eq:S_alphabeta}
\begin{split}
    \mc{S}_{\beta\alpha}^{j}(t')&=[\Delta_{\beta\alpha}^{j}+G_{\beta\alpha}^{j}-iG_{\beta\alpha}^{j}\dot{f}(\omega_{j},t')]S_{\beta\alpha}(t')\\
    &+\bigg[ \hbar(\lambda_{\beta\alpha,\beta\beta}^{j}-\lambda_{\beta\alpha,\alpha\alpha}^{j})\ddot{f}(\omega_{j},t')W_{\beta\alpha}(t')\\
    &+\hbar(\lambda_{\alpha\beta,\beta\beta}^{j}-\lambda_{\alpha\beta,\alpha\alpha}^{j})\ddot{f}(\omega_{j},t')X_{\beta\alpha}(t')\\
    &+i\hbar^{2}\lambda_{\alpha\beta,\beta\alpha}^{j}f^{(3)}(\omega_{j},t')\bigg] \Pi_{\beta\alpha}(t')
\end{split}
\end{equation}
in which the concrete expressions for the time profiles on the right-hand side are given by \eqs{eq:trace_sum_jum}{eq:blocks_S}.

To obtain a continuous expression for the rate of dissipation at site $A$ within the frequency window $[\omega,\omega+d\omega]$ at a specific time, we first identify the $j$ modes associated with site $A$. We then introduce the substitution $\lambda_{\beta\alpha,\gamma\delta}^{j} \to \frac{J^{A}_{\beta\alpha,\gamma\delta}(\omega)}{\omega} d\omega$. With this definition, the rate of dissipation at site $A$ becomes 
\begin{equation}
\mathcal{D}_{A}(\omega,t)\:d\omega = \sum_{\alpha}\sum_{\beta\neq\alpha}\mathcal{J}_{\beta\alpha}^{A}(\omega) P_{\alpha}(t)\: d\omega,
\end{equation}
where $\mathcal{J}_{\beta\alpha}^{A}(\omega)$ is analogous to the expression in \eq{eq:rateconst_diss_int}, but incorporates the aforementioned substitution $\lambda_{\beta\alpha,\gamma\delta}^{j} \to \frac{J^{A}_{\beta\alpha,\gamma\delta}(\omega)}{\omega} d\omega$. 

This formalism shares a structural similarity with our previously developed QME-D method (see eq. (18) and eq. (19) in Ref. \cite{kim2024general1}), but this new approach introduces fundamental distinctions. Specifically, the present formalism is developed in the exciton basis using the system-bath coupling as the perturbation, while our QME-D method operates in the site basis and perturbs the system's electronic coupling. A further distinction lies in the complexity of the final dissipation rate expressions. The current approach, by its construction, generates additional terms dependent on higher-order time derivatives of the bath response function (the final three terms in \eq{eq:S_alphabeta}). These terms, which do not have a counterpart in the QME-D framework, allow for a more detailed description of the dissipative dynamics at the cost of a more computationally demanding implementation.

The accumulated site dissipation at a given time, $\mathcal{E}_{A}(\omega,t)$, can then be obtained as 
\begin{equation}\label{eq:MRT-D-Site}
    \mathcal{E}_{A}(\omega,t) = \int_{0}^{t} \mathcal{D}_{A}(\omega,t') dt'.
\end{equation}
In turn, the total time-dependent dissipation can be obtained as
\begin{equation}\label{eq:MRT-D-Total}
    \mathcal{E}(\omega,t)=\sum\limits_{A=1}^{N}\mathcal{E}_{A}(\omega,t),
\end{equation}
where $N$ is the number of sites.

\begin{figure*}[ht]
    \centering
    \includegraphics[width=0.7\textwidth]{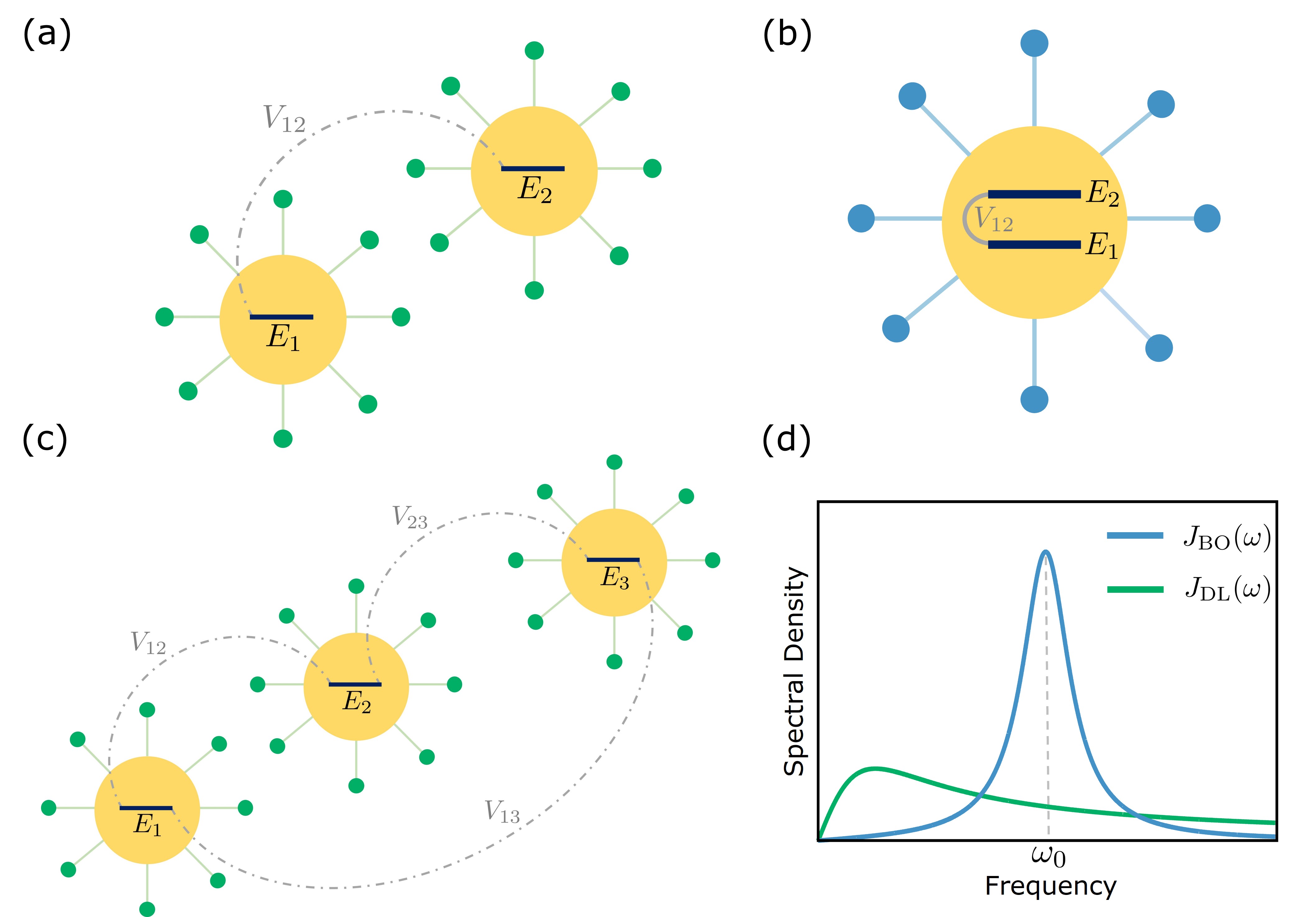}
    \caption{\textbf{Schematic representation of the model systems.} (a) Molecular dimer model where each system state ($E_1$, $E_2$) is connected with its own set of harmonic oscillators (represented by green circles). $V_{12}$ is the coupling connecting the system states. (b) Spin-boson model where the system states are connected to the same bath of harmonic oscillators (represented by blue circles). (c) Same as in (a) but for a molecular trimer model. (d) Shows the functional forms of the Drude-Lorentz spectral density (\eq{eq:DL-SPD-5}, green line) for the molecular dimer/trimer simulations and the Brownian Oscillator spectral density (\eq{eq:BO-SPD}, blue line) for the spin-boson simulations.}
    \label{fig:scheme-models}
\end{figure*}

\section{Results and Discussion} \label{sec:results-5}
To evaluate the numerical accuracy of the proposed theoretical framework to capture dissipation, we will present results from a comprehensive set of simulations and compare them against benchmark data obtained using the Hierarchical Equations of Motion (HEOM-D) method. We focus on dissipation dynamics within representative open quantum system models that feature harmonic bath modes, as detailed in \stn{subsection:hobath}. However, it is important to reiterate that the developed framework maintains its applicability to a broader range of environments, whether harmonic or anharmonic, provided they consist of independent bath degrees of freedom.

The system-bath interactions in these simulations are primarily characterized by the widely used Drude-Lorentz (DL) and Brownian Oscillator (BO) spectral densities. The Drude-Lorentz spectral density, which is often employed to describe the collective low-frequency motions of a solvent environment, is expressed as
\begin{equation}\label{eq:DL-SPD-5}
    J_{\text{DL}}(\omega) = \frac{2 \Lambda}{\pi} \frac{\omega_{c}\omega}{\omega^{2}+\omega_{c}^{2}},
\end{equation}
where $\Lambda$ is the total reorganization energy, which measures the overall strength of the system-bath coupling, and $\omega_{c}$ is the cutoff frequency, which dictates the characteristic relaxation timescale of these bath modes. In turn, the Brownian Oscillator spectral density is typically used for modeling more specific, often higher-frequency, intramolecular vibrational modes of the molecule. Its mathematical form is
\begin{equation}\label{eq:BO-SPD}
    J_{\text{BO}}(\omega) = \frac{2 \Lambda \gamma}{\pi} \frac{2\omega_{0}^{2}\omega}{(\omega^{2}-\omega_{0}^{2})^{2}+4\gamma^{2}\omega^{2}},
\end{equation}
where $\omega_{0}$ is the frequency of the intramolecular vibration and $\gamma$ the damping strength.

The section is structured around four distinct sets of simulations, each designed to probe different aspects and parameter dependencies of the system and bath dynamics:
\begin{itemize}
    \item \textbf{Simulation Set A} focuses on a molecular dimer where each site interacts with its own low-frequency bath components, as described by the Drude-Lorentz spectral density detailed above. Key parameters such as the system site coupling ($V$), temperature ($T$), and the DL cutoff frequency ($\omega_{c}$) are held constant at ($0.25, 1.0, 0.5$), respectively. While we vary the reorganization energy  ($\Lambda$) across values of ($0.05, 0.2, 1.0, 2.0$) and the energy gap ($\Delta E$) among ($0.5, 1.0, 2.0$). For all 12 conditions within this set, the initial excitation is placed at the upper exciton state.
    \item \textbf{Simulation Set B} utilizes the same molecular dimer model as Simulation Set A, where each site interacts with a bath described by the Drude-Lorentz spectral density detailed above, but investigates the effects of varying temperature. Key parameters such as the system site coupling ($V$), reorganization energy ($\Lambda$), and the DL cutoff frequency ($\omega_{c}$) are held constant at (0.25,0.2,0.5), respectively. While we vary the temperature ($T$) across values of ($0.25,0.5,1.0$) and the energy gap ($\Delta E$) among ($0.5,1.0,2.0$). For all 9 conditions within this set, the initial excitation is placed at the upper exciton state.
    \item \textbf{Simulation Set C}  examines a spin-boson model, representing a two-level system coupled to a single bath. In this case, the Brownian Oscillator spectral density detailed above characterizes the system-bath interaction. Key parameters such as the system site coupling ($V$), temperature ($T$), energy gap ($\Delta E$), and the BO characteristic frequency ($\omega_{0}$) are held constant at ($0.25,1.0,2.0,2.062$), respectively. In contrast, we vary the reorganization energy ($\Lambda$) across values of ($0.05,0.25,1.0$) and the BO damping strength ($\gamma$) among ($0.05,0.25,1.0$). For all 9 conditions within this set, the initial excitation is placed at the upper exciton state.
    \item \textbf{Simulation Set D} extends the analysis to a molecular trimer where each site interacts with low-frequency components of the bath, described by the Drude-Lorentz spectral density detailed above. For this set, several key parameters are held constant. The temperature ($T$) is maintained at 1.0, and the DL cutoff frequency ($\omega_{c}$) is 0.5. The inter-site energy gaps are fixed with $\Delta E_{12}=1.0$, $\Delta E_{13}=2.0$, and $\Delta E_{23}=1.0$. Specific site couplings are also set at $V_{13}=0.25$ and $V_{23}=0.0$. In turn, we vary the reorganization energy ($\Lambda$) across values of ($0.1,0.5$) and the site coupling $V_{12}$ among ($0.25,0.5,1.0$). Furthermore, three distinct initial excitation schemes are employed. The first scheme involves placing the initial excitation at the middle site energy state (site 2). This setup compares QME-D\cite{kim2024general1,kim2024general2} results against HEOM benchmarks to demonstrate the regime of applicability of the QME-D theory. The second scheme utilizes an incoherent mixture of states on an exciton basis, representing an initial excitation localized at site 2, aiming to show how our framework can overcome certain limitations inherent in the QME-D approach. The third scheme places the initial excitation directly into the middle exciton state. There is a total of 18 distinct simulation conditions for this set.
\end{itemize}

A schematic of the model systems and spectral densities used in this section is presented in \fig{fig:scheme-models}.

To incorporate non-Markovian effects, we implemented the time scale separation method (TSS)\cite{berkelbach2012reduced,montoya2015extending}. This method separates the spectral density into slow and fast components, with only the fast component directly influencing the system dynamics and the slow components treated as a source of static disorder. The spectral density separation is formally achieved by defining
\begin{equation}
\begin{split}
J_{\text{slow}}(\omega)&= S(\omega,\omega^{*})J(\omega)\\
J_{\text{fast}}(\omega)&=[1-S(\omega,\omega^{*})]J(\omega)
\end{split}
\end{equation}
where $S(\omega,\omega^{*})$ is the splitting function given by
\begin{equation}\label{eq:splitting}
S(\omega,\omega^{*})=\left\{ \begin{array}{ccc} \eta[1-(\omega/\omega^{*})^{2}]^{2}, & & \omega<\omega^{*}\\ 0, & & \omega\geq\omega^{*} \end{array}\right.
\end{equation}
and $\omega^{*}$ is the cutoff frequency. 

As the Simulation Sets described here are consistent with Modified Redfield Theory, we will now refer to our method for computing dissipation as MRT-D.

\begin{table*}[btp] 
    \centering
    \setlength{\tabcolsep}{4pt}
    \small
    \caption{Simulation parameters for the molecular dimer model. The reorganization energy ($\Lambda$) and temperature ($T$) define the system–bath interaction, while the remaining parameters specify the HEOM procedure. Each of the six conditions listed was combined with three different energy gap values ($\Delta$ E = 0.5,1.0,2.0), resulting in 18 distinct simulations.}
    \begin{tabular*}{\textwidth}{@{\extracolsep{\fill}} lcccccc}
    \hline \hline
         Simulation Condition&  (i)& (ii) & (iii) & (iv) & (v) &(vi) \\ \hline
         Reorganization Energy ($\Lambda$)& 0.05 & 0.2 &  1.0 & 2.0 & 0.2 & 0.2\\
         Temperature ($T$)& 1.0  & 1.0 & 1.0 & 1.0 & 0.5 & 0.25 \\
         Maximum time step ($\Delta t_{\ti{max}}$)& 0.02 & 0.1 & 0.05 & 0.05 & 0.1 & 0.1\\
         Number of hierarchy tiers ($N_{\text{hier}}$)& 4 & 7 & 10 &13  & 7 & 7\\
         Number of Matsubara terms ($N_{\text{Matsu}}$)& 30 & 30 & 30 & 30 & 100 & 100\\
         H-R factor of the probe mode ($s_{\ti{pb}}$)& $2. 10^{-6}$ & $1.10^ {-5}$ & $1. 10^{-5}$ & $1. 10^{-5}$ & $1. 10^{-5}$ &$1. 10^{-5}$ \\
         \hline \hline
    \end{tabular*}
    \label{tab:HEOM-D-param}
\end{table*}
\subsection{Molecular dimer}
\subsubsection{Simulation details} \label{sec:SD-MD}
The simulations for the molecular dimer model, comprising Simulation Sets A and B as defined previously, were conducted using Planck atomic units ($\hbar=k_{\ti{B}}=1$). For all scenarios within these sets, the initial excitation was placed at the upper exciton state, and the energy gaps ($\Delta E$) was varied across the values ($0.5,1.0,2.0$).

In the MRT-D computations, each Drude-Lorentz bath spectral density (BSD) associated with the dimer sites was discretized into 2000 harmonic oscillator modes. This discretization followed the scheme detailed in \app{sec:app_discretization}. An upper frequency limit of $\omega_{\ti{max}}=15$ was set, which recovered $97.9\%$ of the analytical BSD's reorganization energy. Time integrals for determining exciton population rates, \eq{eq:pop-rate}, and dissipation rate constants, \eq{eq:rateconst_diss_int}, were evaluated using the trapezoidal method with an integration grid size of 0.02 and an upper integration limit of $5\times 10^{3}$. The coupled rate equations for exciton populations, \eq{eq:rateeqn_pop}, were then propagated using a fourth-order Runge-Kutta algorithm with a time step of 0.02.\cite{fehlberg1969low}

The Time Scale Separation (TSS) method was incorporated into MRT-D calculations to explore the influence of non-Markovian memory effects on dimer dynamics. In these specific dimer simulations, the splitting function, \eq{eq:splitting}, was defined by setting  $\eta=0.99$ and $\omega^{*}=0.05$. The final results were obtained by averaging over many trajectories. Specifically, $10^{4}$ trajectories were used for the conditions involving a reorganization energy $\Lambda=0.05$ (part of Simulation Set A), while $10^{3}$ trajectories were used for all other conditions within Simulation Sets A and B. The number of trajectories was chosen to ensure numerical convergence.

For comparison, numerically exact benchmarks for the dissipation dynamics were established using the Hierarchical Equations of Motion (HEOM) method. This was implemented using the HEOM-D strategy for monitoring bath components developed by Kim\cite{kim_extracting_2022}, along with an efficient low-temperature correction scheme recently reported.\cite{fay2022simple} Key HEOM parameters for each simulation condition, including the hierarchy depth ($N_{\text{hier}}$), the number of Matsubara terms ($N_{\text{Matsu}}$), and the Huang-Rhys (H-R) factor for the probe mode ($s_{\ti{pb}}$), are detailed in Table \ref{tab:HEOM-D-param}. 

The frequency of the HEOM-D probe mode was scanned from 0.1 to 3.0 in steps of 0.05. For probe frequencies $\omega\geq0.2$, the number of vibrational quantum states describing the probe was chosen to ensure the initial bath density represented $99.9\%$ of the total Boltzmann population. This threshold was relaxed to $99.0\%$ for $\omega < 0.2$ to mitigate the rapidly increasing computational burden at lower frequencies. The approach to steady-state ($t\to\infty$) was practically handled by defining a finite simulation time, $t_{\ti{sim}}$, for each condition. This time was determined by visually inspecting the convergence of excitonic population dynamics. The system reduced density matrix (RDM) and the associated auxiliary density matrices (ADMs) were propagated using an adaptive RKF45 integrator. The time step was dynamically adjusted based on the deviation of the RDM's trace from unity. To further ensure numerical stability, especially near steady state, the integration time step was not permitted to exceed a predefined maximum, $\Delta t_{\ti{max}}$.

\subsubsection{Electronic dynamics}
Accurate exciton populations are a fundamental prerequisite to reliably capturing energy dissipation dynamics. For this reason, we first examine the fidelity of the population dynamics predicted by MRT. We evaluate this accuracy by describing the time evolution of exciton populations within the molecular dimer and benchmarking MRT's predictions against numerically exact HEOM results. This comparison is illustrated in Figure 1 for Simulation Set A and Figure 2 for Simulation Set B, which display the time-dependent population inversion $\langle \hat{\sigma}_{z} (t)\rangle=P_{\alpha}(t)-P_{\beta}(t)$ where $P_{\alpha}$ and $P_{\beta}$ are the lower and higher excitonic states populations, respectively. Both figures compare MRT (cyan line) with HEOM (pink dashed line).

\begin{figure*}[ht]
    \centering
    \includegraphics[width=0.9\textwidth]{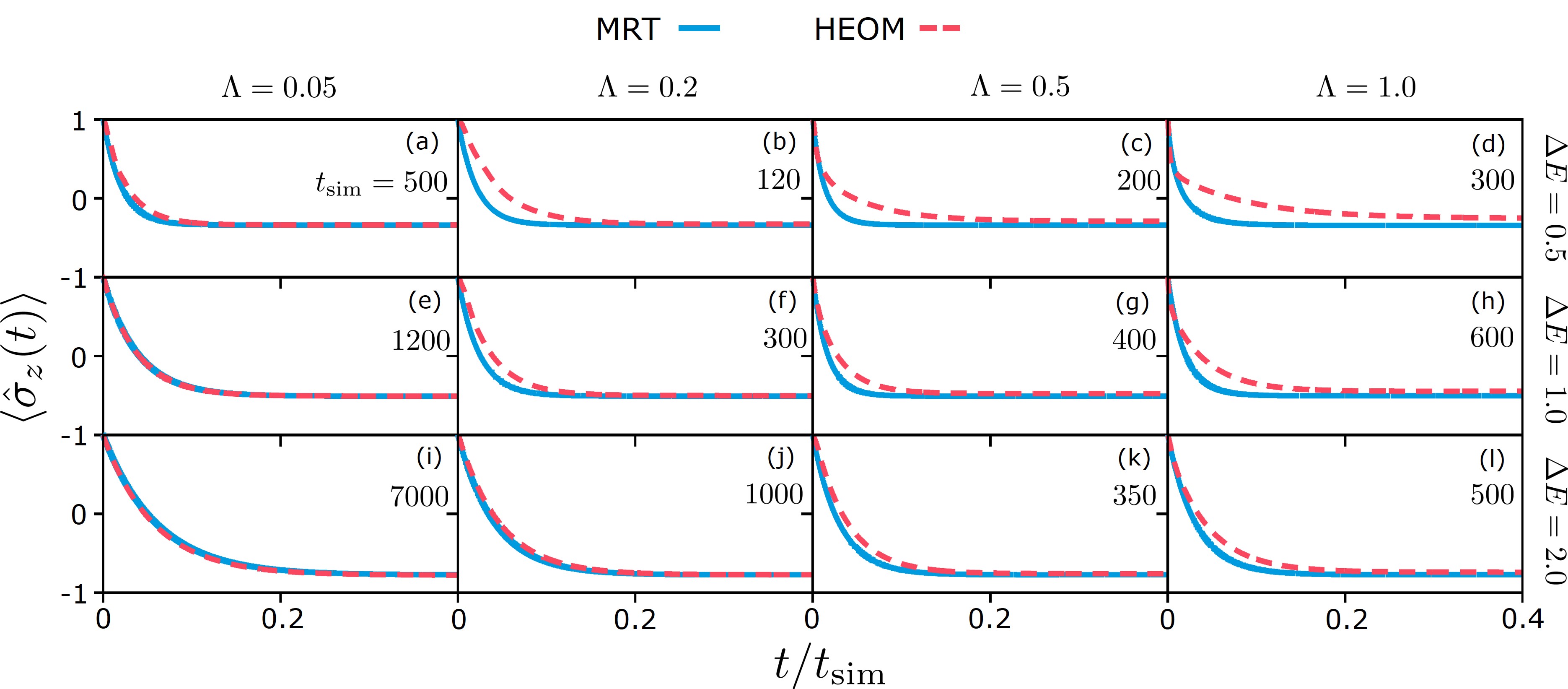}
    \caption{\textbf{ Time-dependent population inversion, $\langle \hat{\sigma}_{z} (t)\rangle=P_{\alpha}(t)-P_{\beta}(t)$, for the molecular dimer in Simulation Set A.} Results from MRT (cyan line) are compared with HEOM benchmarks (pink dashed line). Panels illustrate the dynamics for different reorganization energies ($\Lambda = \{0.05,0.2,0.5,1.0\}$) and energy gaps ($\Delta E= \{0.5,1.0,2.0\}$), with fixed parameters $T=1.0$, $V=0.25$, and a Drude-Lorentz cutoff frequency $\omega_{c}=0.5$.}
    \label{fig:pop-SS1}
\end{figure*}

\Fig{fig:pop-SS1} (Simulation Set A) shows that MRT performs well across a significant range of conditions, yielding good agreement with HEOM. While, in general, MRT provides good predictions, some deviations from HEOM benchmarks emerge under specific conditions, notably highlighted in panels (b)-(d) and (h). These cases exemplify situations where the reorganization energy $\Lambda \geq 0.2$ and the energy gap $\Delta E \leq 1.0$, leading MRT to overestimate the rate of population transfer. The challenge for MRT is most apparent when strong coupling ($\Lambda =1.0$) combines with a small energy gap, $\Delta E = 0.5$, as seen in \fig{fig:pop-SS1}d, where the quantitative accuracy in capturing the precise steady state and transfer rate is more limited. This behavior is attributed to MRT's underlying assumption of an intermediate system-bath coupling strength relative to the system energy gap.

Conversely, MRT's accuracy significantly improves as the energy gap increases. This enhanced performance arises because a larger energy gap promotes more localized exciton states. Consequently, the dynamics become predominantly diffusive and characterized by an incoherent exponential decay. This characteristic behavior becomes clear by solving the coupled differential equations for state populations, \eq{eq:rate_prim}, which leads to an analytical expression for the population inversion given by
\begin{equation}
    \langle \hat{\sigma}_{z} (t)\rangle = \langle \hat{\sigma}_{z} (\infty)\rangle +[\langle \hat{\sigma}_{z} (0)\rangle-\langle \hat{\sigma}_{z} (\infty)\rangle]\exp{(K_{\beta \alpha}-K_{\alpha \beta})t}.
\end{equation}
    
\begin{figure}[ht]
    \centering
    \includegraphics[width=1.0\columnwidth]{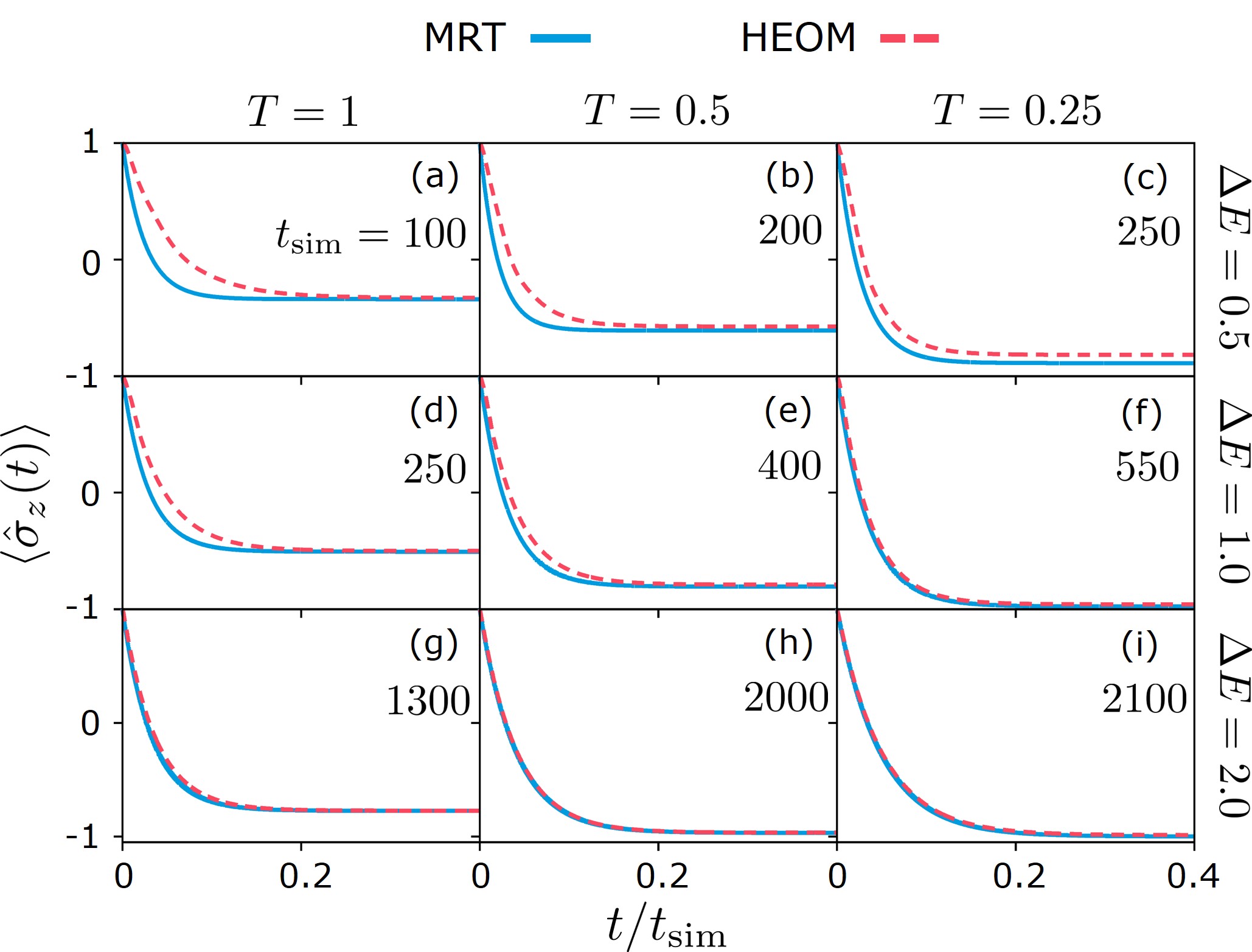}
    \caption{\textbf{ Time-dependent population inversion, $\langle \hat{\sigma}_{z} (t)\rangle=P_{\alpha}(t)-P_{\beta}(t)$, for the molecular dimer in Simulation Set B.} Results from MRT (cyan line) are compared with HEOM benchmarks (pink dashed line). Panels illustrate the dynamics for different temperatures ($T = \{0.25,0.5,1.0\}$) and energy gaps ($\Delta E = \{0.5,1.0,2.0\}$), with fixed parameters $\Lambda=0.2$, $V=0.25$, and a Drude-Lorentz cutoff frequency $\omega_{c}=0.5$.}
    \label{fig:pop-SS2}
\end{figure}

\Fig{fig:pop-SS2} presents the results for Simulation Set B, which are similar in trends to those in \Fig{fig:pop-SS1}, as MRT accuracy improves with an increasing energy gap ($\Delta E$), and slightly overestimate population transfer rates as the energy gap is reduced and the temperature increased. This temperature deviation is attributed to greater thermal fluctuations induced by the system-bath interaction.

\subsubsection{Dissipation dynamics}
Having established MRT's performance for population dynamics in the preceding section, we now consider evaluating the dissipation predicted by MRT-D. We will compare these predictions against HEOM-D benchmarks for Simulation Sets A and B. We will focus on total dissipation to maintain visual clarity in the analysis. A more detailed examination of site-specific dissipation contributions will be explored in the subsequent discussion of the molecular trimer system (\stn{sn:trimer}).

\begin{figure*}[ht]
    \centering
    \includegraphics[width=0.9\textwidth]{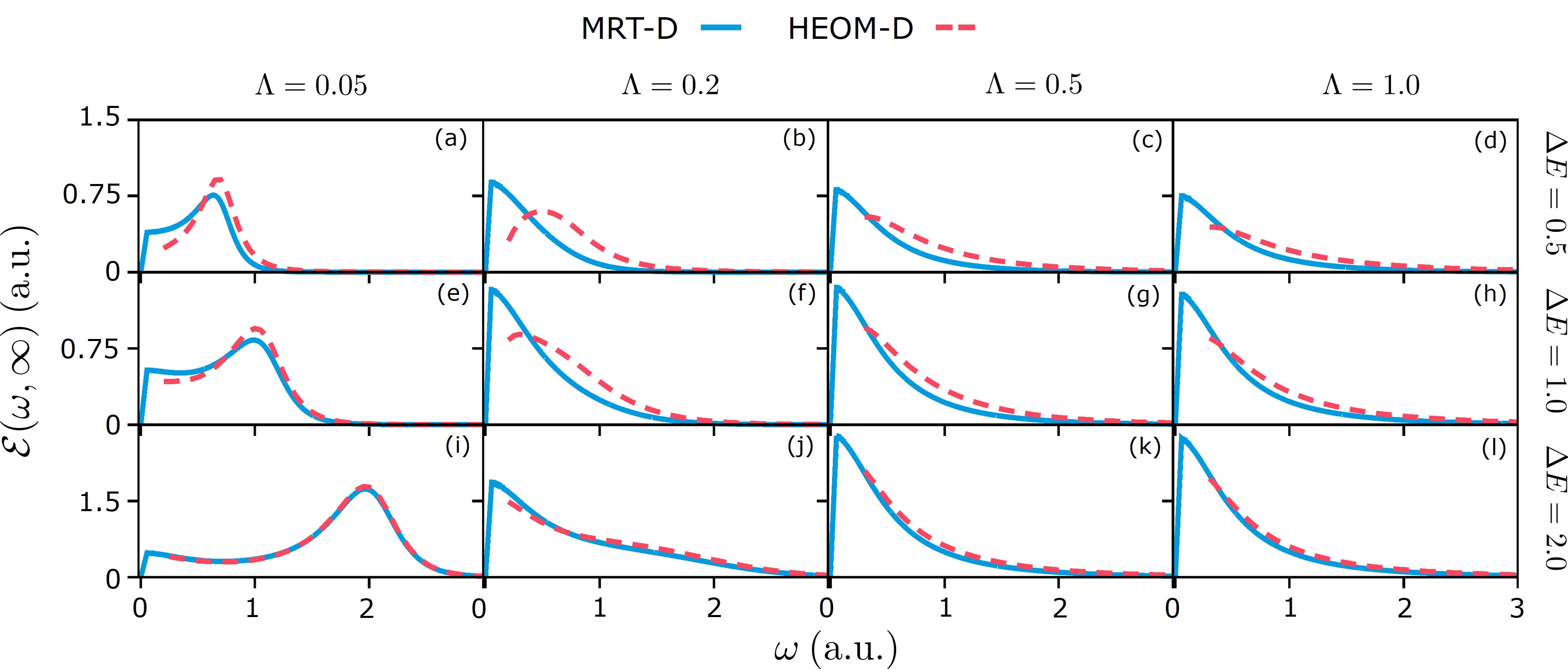}
    \caption{\textbf{ Total steady-state dissipation density, $\mathcal{E}(\omega,\infty)$
     for the molecular dimer in Simulation Set A.} Results from MRT-D (cyan line) are compared with HEOM-D benchmarks (pink dashed line). Panels illustrate the dynamics for different reorganization energies ($\Lambda = \{0.05,0.2,0.5,1.0\}$) and energy gaps ($\Delta E= \{0.5,1.0,2.0\}$), with fixed parameters $T=1.0$, $V=0.25$, and a Drude-Lorentz cutoff frequency $\omega_{c}=0.5$.}
    \label{fig:dis-SS1}
\end{figure*}

The frequency-resolved dissipation can be accessed through the accumulated dissipation density, $\mathcal{E}(\omega,t)$, as defined in \eq{eq:MRT-D-Total}. \Fig{fig:dis-SS1}  illustrates the steady-state cumulative dissipation, $\mathcal{E}(\omega,\infty)$ for Simulation Set A, comparing the results obtained from MRT-D (cyan line) and HEOM-D (pink dashed line). Consistent with the trends observed for population dynamics presented in \fig{fig:pop-SS1}, the accuracy of MRT-D in predicting dissipation improves with decreasing reorganization energy ($\Lambda$) and increasing energy gap ($\Delta E$). It is particularly noteworthy that even for the challenging condition of $\Lambda=1.0$ and $\Delta E=0.5$ (\fig{fig:dis-SS1}d), where MRT showed quantitative deviations in population dynamics, the MRT-D framework still yields a qualitatively good description of the dissipation spectrum.

Examining the dissipation mechanisms revealed in \fig{fig:dis-SS1}, we observe distinct behaviors dependent on the system-bath coupling strength. With a relatively small reorganization energy of $\Lambda=0.05$, a substantial portion of the energy dissipation occurs through a channel centered around $\hbar\omega=\Delta E$. This feature can be attributed to vibronic resonance, where energy is efficiently transferred to quasi-resonant bath modes with the excitonic energy difference. As the reorganization energy $\Lambda$ is increased, the contribution of this vibronic resonance channel gradually diminishes, and the dissipation becomes increasingly concentrated at lower frequencies, approaching $\omega=0$. This shift indicates that stronger coupling promotes dissipation into slower, collective bath motions.

\begin{figure}[ht]
    \centering
    \includegraphics[width=1.0\columnwidth]{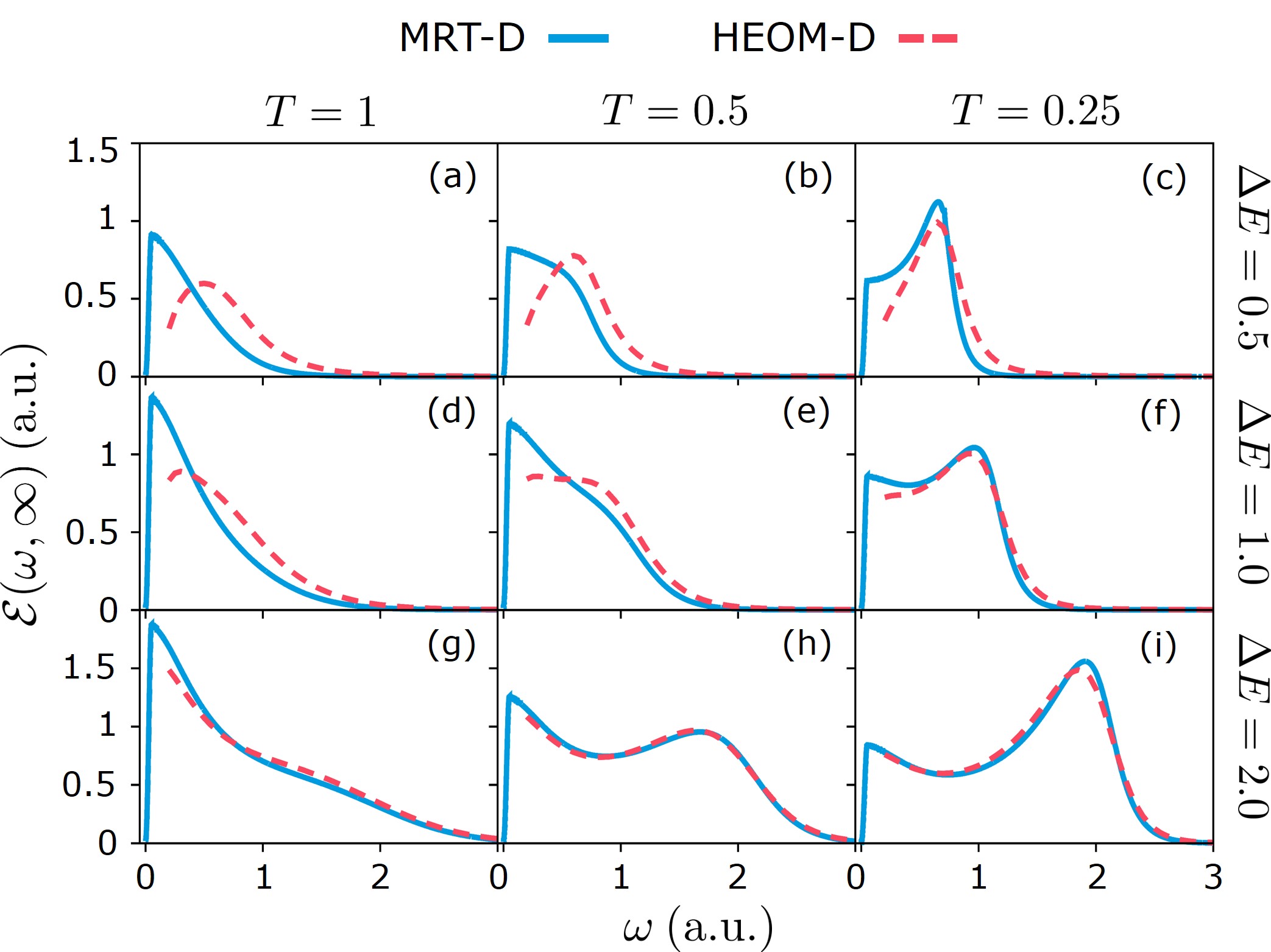}
    \caption{\textbf{ Total steady-state dissipation density, $\mathcal{E}(\omega,\infty)$
     for the molecular dimer in Simulation Set B.} Results from MRT (cyan line) are compared with HEOM benchmarks (pink dashed line). Panels illustrate the dynamics for different temperatures ($T = \{0.25,0.5,1.0\}$) and energy gaps ($\Delta E = \{0.5,1.0,2.0\}$), with fixed parameters $\Lambda=0.2$, $V=0.25$, and a Drude-Lorentz cutoff frequency $\omega_{c}=0.5$.}
    \label{fig:dis-SS2}
\end{figure}

In turn, \fig{fig:dis-SS2} shows the influence of temperature on the steady-state total accumulated dissipation, $\mathcal{E}(\omega,\infty)$, for Simulation Set B, comparing MRT-D and HEOM-D calculations. MRT-D accuracy generally increases with a larger energy gap ($\Delta E$), and it successfully captures the correct qualitative trends across the temperature series. Notably, lowering the temperature ( from $T=1.0$ to $T=0.25$) enhances the prominence of the vibronic resonance channel in the dissipation spectrum. This enhancement results from the reduction in thermal fluctuations induced by the system-bath interaction at lower temperatures, which allows the more specific resonant energy transfer processes to become more dominant.

\subsection{Spin-boson model with Brownian oscillator bath}
\subsubsection{Simulation details} \label{sim-det}
The simulations for the Spin-boson model correspond to Simulation Set C. These calculations were performed using Planck atomic units ($\hbar=k_{\text{B}}=1$). The model features a two-level system coupled to a single bath, which is characterized by a Brownian oscillator spectral density, as defined in \eq{eq:BO-SPD}. Key parameters such as the system site coupling ($V=0.25$), temperature ($T=1.0$), energy gap ($\Delta E=2.0$), and the BO characteristic frequency ($\omega_{0}=2.062$) were held constant. We varied the reorganization energy ($\Lambda$) across values of \{$0.05, 0.25, 1.0\}$ 
and the BO damping strength ($\gamma$) among $\{0.05, 0.25, 1.0\}$. For all 9 conditions within this set, the initial excitation was placed at the upper exciton state.

The Brownian oscillator bath spectral density was discretized into 10000 harmonic oscillator modes for the MRT-D computations. The discretization scheme follows the procedure described in \app{sec:app_discretization}. In turn, we set $\omega_{0}=2.062$ as the center of the Brownian spectral density. Time integrals for determining exciton population rates, \eq{eq:pop-rate}, and dissipation rate constants, \eq{eq:rateconst_diss_int}, were evaluated using the trapezoidal method with an integration grid size of $0.02$ and an upper integration limit of $5\times10^{3}$. The coupled rate equations for exciton populations (\eq{eq:rateeqn_pop}) were propagated using a fourth-order Runge-Kutta algorithm with a time step of $0.02$.

The Time Scale Separation method was incorporated into the MRT-D calculations. For the time-scale separation, the splitting function (\eq{eq:splitting}) was defined by setting the cutoff frequency $\omega^{*}=0.05$, and the parameter $\eta$ was reduced from 0.99 (as used in molecular dimer simulations ) to 0.6. This reduction was necessary due to the increased difficulty of achieving detailed balance conditions with the Brownian oscillator bath. To ensure numerical convergence, the number of individual noise trajectories averaged to obtain final results was kept at $10^{4}$.

For the HEOM and HEOM-D simulations, we implemented the Brownian oscillator BSD based on the strategy for monitoring bath components developed by Kim\cite{kim_extracting_2022}, along with an efficient low-temperature correction scheme recently reported.\cite{fay2022simple}
The HEOM-D parameters used for each simulation condition can be found in Table \ref{tab:HEOM-D-param}. 

 \begin{table}[ht] 
    \centering
    \setlength{\tabcolsep}{\fill} 
    \small
    \caption{Simulation parameters for the spin-boson model. The reorganization energy ($\Lambda$) defines the system–bath interaction. The remaining parameters specify the HEOM procedure. Each of the three conditions listed was combined with three different damping values ($\gamma$ = $\{0.5,1.0,2.0\}$), resulting in 9 distinct simulations.}
    \begin{tabular*}{\columnwidth}{@{\extracolsep{\fill}}l c c c}
    \hline \hline
         Simulation Condition&  (i)& (ii) & (iii) \\ \hline
         Reorganization Energy ($\Lambda$)& 0.05 & 0.25 &  1.0 \\
         Maximum time step ($\Delta t_{\ti{max}}$)& 0.01 & 0.05 & 0.05\\
         Number of hierarchy tiers ($N_{\text{hier}}$)& 5 & 7 & 12 \\
         Number of Matsubara terms ($N_{\text{Matsu}}$)& 10 & 15 & 25\\
         H-R factor of the probe mode ($s_{\ti{pb}}$)& $2. 10^{-6}$ & $1.10^ {-5}$ & $1. 10^{-5}$  \\
         \hline \hline
    \end{tabular*}
    \label{tab:HEOM-D-param-BO}
\end{table}
\subsubsection{Electronic and dissipation dynamics}
\begin{figure}[ht]
    \centering
\includegraphics[width=1.0\columnwidth]{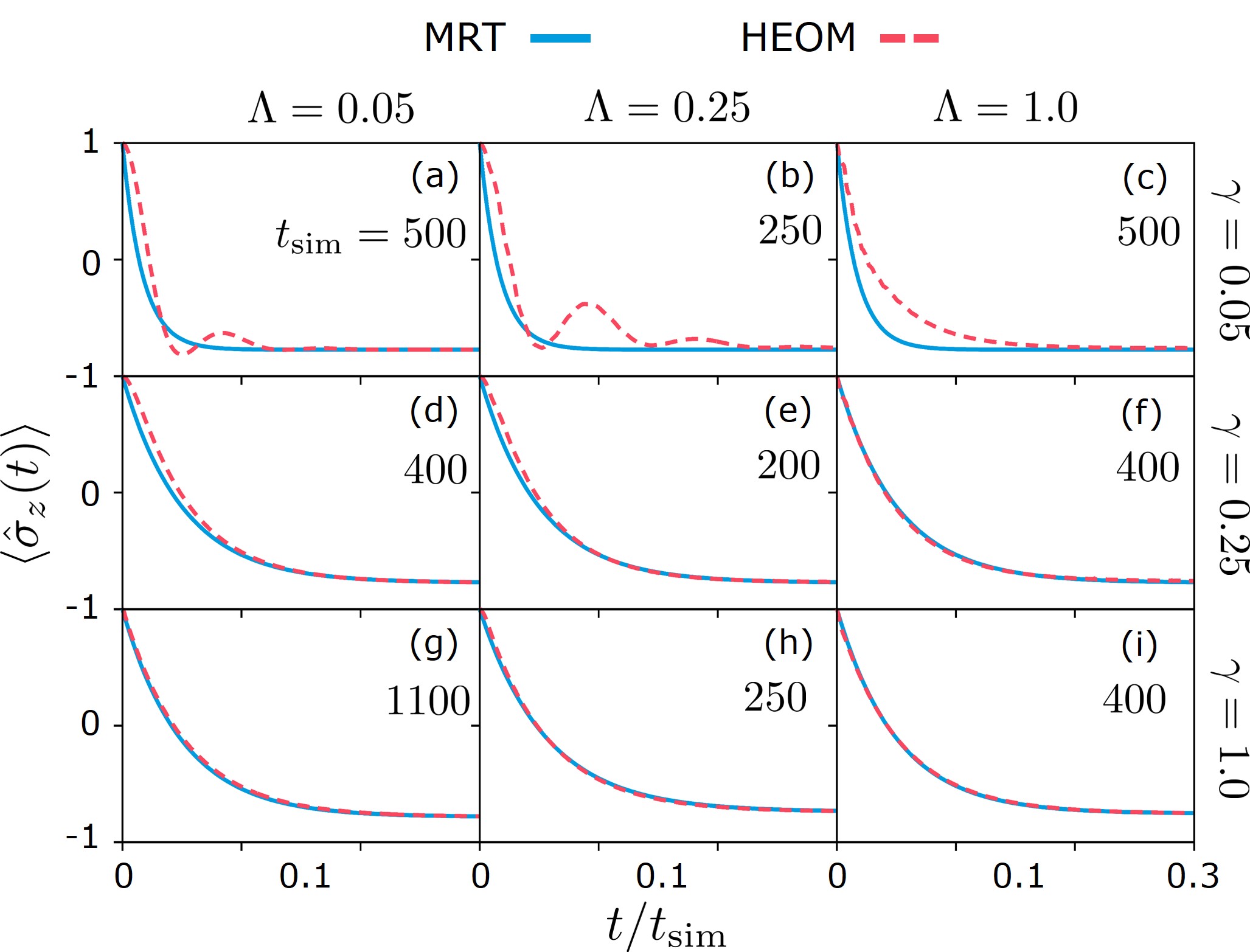}
    \caption{\textbf{ Time-dependent population inversion, $\langle \hat{\sigma}_{z} (t)\rangle=P_{\alpha}(t)-P_{\beta}(t)$, for the spin-boson model in Simulation Set C.} Results from MRT (cyan line) are compared with HEOM benchmarks (pink dashed line). Panels illustrate the dynamics for different reorganization energies ($\Lambda = \{0.05,0.25,1.0\}$) and damping factors ($\gamma= \{0.05,0.25,1.0\}$). All other parameters are as specified in \ref{sim-det}.}
    \label{fig:pop-BO}
\end{figure}

\Fig{fig:pop-BO} presents the time evolution of the population inversion, $\langle \hat{\sigma}_{z}(t)\rangle$, for the nine simulation conditions in Table \ref{tab:HEOM-D-param-BO}, revealing the accuracy of the MRT when benchmarked against HEOM results. For a small damping strength, such as $\gamma=0.05$, \fig{fig:pop-BO}a-c shows that MRT finds it challenging to describe the highly non-Markovian character of the bath dynamics. This difficulty arises from the underdamped nature of the bath, which leads to persistent memory effects. However, as the damping strength ($\gamma$) increases, the agreement between MRT and HEOM benchmarks significantly improves. MRT provides a nearly quantitative match to the benchmark electronic dynamics for larger $\gamma$ values (e.g., 0.25 and 1.0), showing its utility when the bath becomes more dissipative and its memory effects are shortened.

\begin{figure}[ht]
    \centering
\includegraphics[width=1.0\columnwidth]{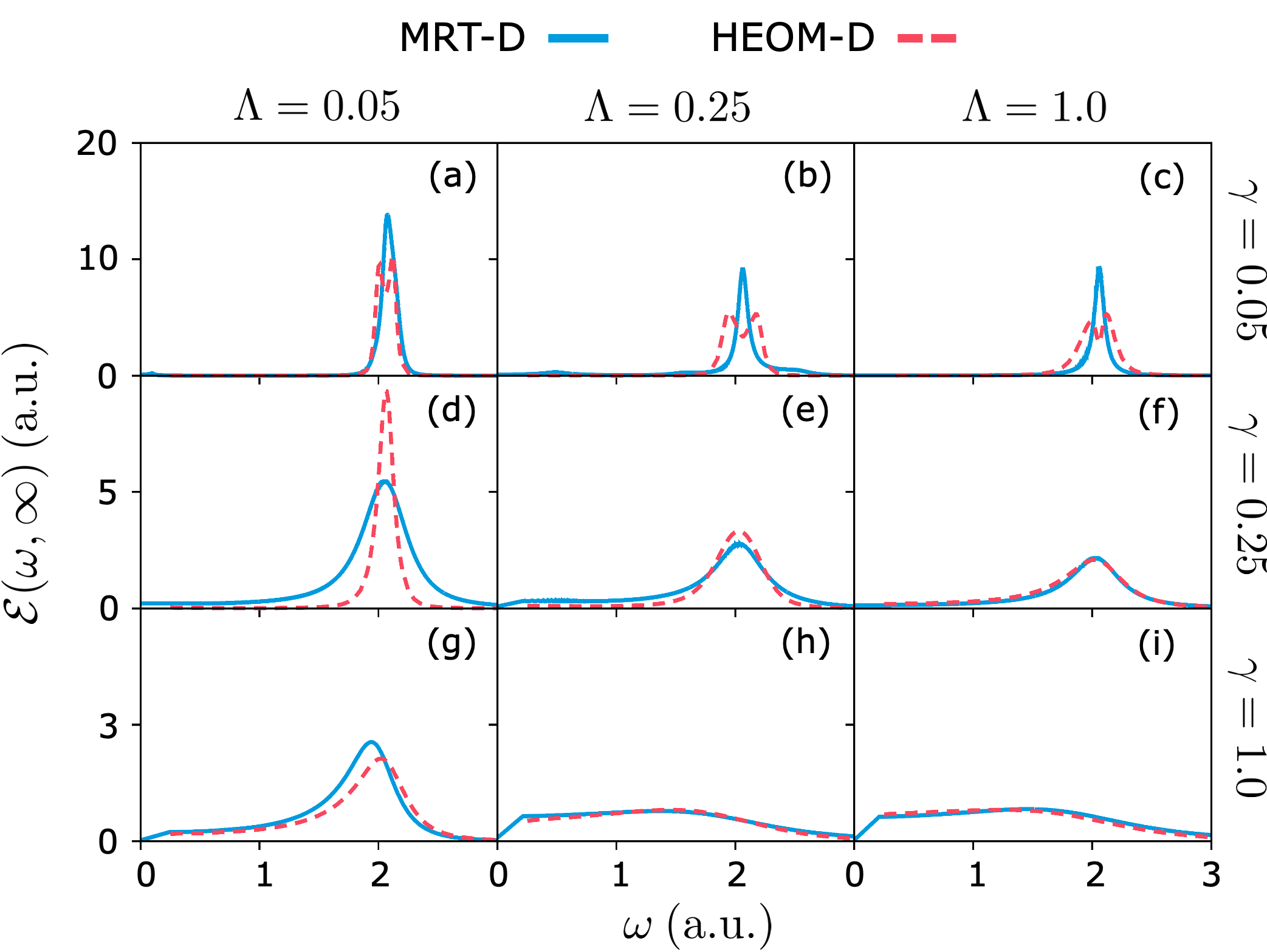}
    \caption{\textbf{ Total steady-state dissipation density, $\mathcal{E}(\omega,\infty)$
     for the molecular dimer in Simulation Set C.} Results from MRT (cyan line) are compared with HEOM benchmarks (pink dashed line). Panels illustrate the dynamics for different reorganization energies ($\Lambda = \{0.05,0.25,1.0\}$) and damping factors ($\gamma= \{0.05,0.25,1.0\}$). All other parameters are as specified in \ref{sim-det}.}
    \label{fig:dis-BO}
\end{figure}

In \fig{fig:dis-BO} we present the steady-state accumulated dissipation density, $\mathcal{E}(\omega,\infty)$, in MRT-D and HEOM-D. The results in \fig{fig:dis-BO} show that for small ($\gamma=0.05$) and intermediate ($\gamma=0.25$) damping strengths, most of the energy dissipation occurs through a resonant channel around $\omega \approx 2.0$. This frequency corresponds closely to both the system's energy gap ($\Delta E=2.0$) and the characteristic frequency of the Brownian oscillator ($\omega_{0}=2.062$). An interesting feature observed at a very small damping strength ($\gamma=0.05$), \fig{fig:dis-BO}a-c, is that the dissipation spectrum in HEOM-D does not form a single peak, which would mirror the shape of the Brownian oscillator spectral density itself. Instead, it presents as a pair of closely lying peaks. Such a spectral structure arises from the strong coherent interaction between the upper subsystem state and the first excited state of the underdamped bath mode, a phenomenon analogous to the formation of polaritonic states. However, this distinct peak-splitting behavior diminishes and eventually disappears as the resonance effect is diluted due to increased damping strength.

For all conditions displayed in \fig{fig:dis-BO}, the MRT-D framework qualitatively reproduces the general features of the dissipation spectra obtained from HEOM-D calculations. The predictive capability of MRT-D improves with increasing damping strength, which enhances the accuracy of the Markov approximation. Interestingly,  at a small reorganization energy ($\Lambda=0.05$), MRT-D deviates from the HEOM-D calculations. We attribute this behavior to the absence of low-frequency components in the bath, which results in long-time bath memory, which decreases the accuracy of the Markov approximation.

\subsection{Molecular trimer}\label{sn:trimer}
\subsubsection{Simulation details}
The simulations for the molecular trimer model (Simulation Set D) were performed using Planck atomic units ($\hbar = k_{\text{B}}=1$). In this model, each site of the trimer interacts with its own low-frequency bath components, characterized by a Drude-Lorentz spectral density. Key fixed parameters for these simulations include a temperature $T=1.0$  and a DL cutoff frequency $\omega_{c}=0.5$. The inter-site energy gaps were set to $\Delta E_{12}=1.0$, $\Delta E_{13}=2.0$, and $\Delta E_{23}=1.0$. Specific site couplings are also set at $V_{13}=0.25$ and $V_{23}=0.0$. The simulations explored variations in the reorganization energy ($\Lambda=\{0.1,0.5\}$) and the site coupling $V_{12}=\{0.25,0.5,1.0\})$. A total of 18 distinct simulation conditions were examined, encompassing three different initial excitation schemes. The first scheme involved placing the initial excitation at the middle site energy state to compare QME-D results against HEOM-D benchmarks. The second scheme utilized an incoherent mixture of states on an exciton basis, representing an initial excitation localized at site 2. This aims to show how MRT-D can overcome certain limitations inherent in the QME-D approach. The third scheme placed the initial excitation directly into the middle exciton state.

The computational details for QME, MRT, QME-D, and MRT-D, including the discretization of the Drude-Lorentz bath spectral density, the parameters for the Time Scale Separation method, the numerical evaluation of time integrals for rates, and the propagation of population equations, were identical to those described in \stn{sec:SD-MD}. Similarly, the HEOM and HEOM-D benchmark calculations, including the HEOM-D strategy for monitoring bath components, the low-temperature correction scheme, the setup for the probe mode scan, the determination of simulation time for steady-state, and the RDM/ADM propagation techniques, followed the procedures outlined in \stn{sec:SD-MD} and parameters detailed in Table \ref{tab:HEOM-D-param}. 

\subsubsection{Electronic dynamics}
In this section, we investigate the electronic population dynamics of the molecular trimer (Simulation Set D). The primary goal is to demonstrate that MRT can accurately capture dynamics under conditions where the QME approach fails. As the QME and MRT frameworks are formulated in different bases-- the site and exciton bases, respectively-- the HEOM benchmark results are presented in the appropriate basis for each comparison.

\begin{figure}[ht]
    \centering    \includegraphics[width=1.0\columnwidth]{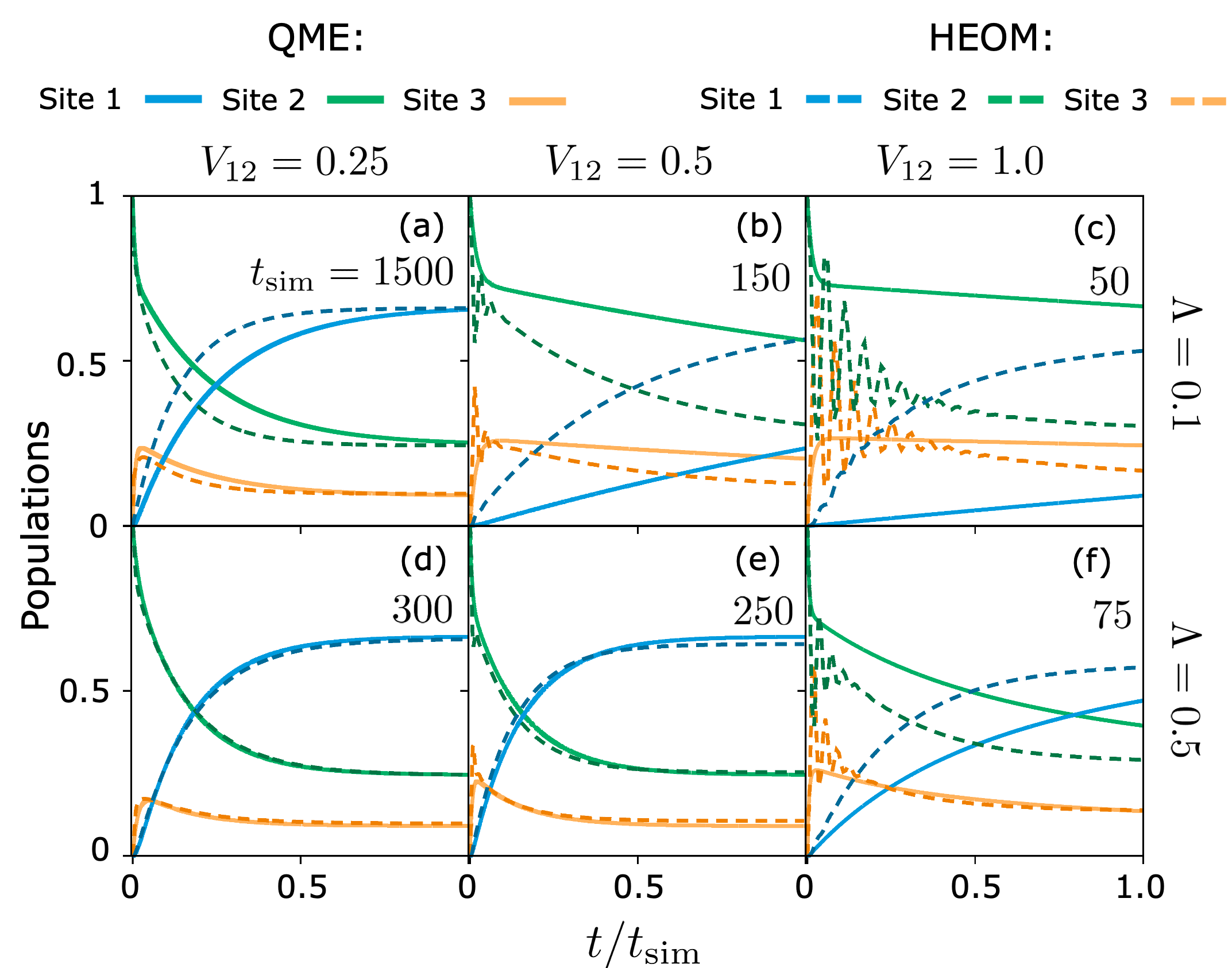}
    \caption{\textbf{Population dynamics for the molecular trimer (Simulation Set D) with initial excitation localized at the middle site energy state.}  Results from QME calculations (full lines) are compared against HEOM (dashed lines). Site populations are color-coded: site 1 (blue), site 2 (green), and site 3 (orange). The fixed simulation parameters are $T=1.0$, $\omega_{c}=0.5$, $\Delta E_{12}=1.0$, $\Delta E_{13}=2.0$,  $\Delta E_{23}=1.0$,  $V_{13}=0.25$ and $V_{23}=0.0$.}
    \label{fig:pop-3a}
\end{figure}

First, to establish the limitations of the QME approach, we initialized the system with the excitation localized on the middle site energy state. The population dynamics are presented in \fig{fig:pop-3a}, where full lines represent QME calculations and dashed lines depict HEOM benchmarks (site 1: blue, site 2: green, site 3: orange). As evident from \fig{fig:pop-3a}, the QME accurately captures the population dynamics only under particular conditions of small electronic coupling ($V_{12}=0.25$) and large reorganization energy ($\Lambda=0.5$), as shown in \fig{fig:pop-3a}d. For the majority of other conditions tested (\fig{fig:pop-3a}a-c, and \fig{fig:pop-3a}f), QME fails to reproduce the correct dynamical behavior. Even in cases like \fig{fig:pop-3a}e, where QME might eventually reach the correct steady-state populations, it fails to describe the short-time dynamics accurately.

To address these limitations, we examine the MRT performance against HEOM. For this, we use an incoherent exciton initial condition, which corresponds to an initial excitation localized at site 2. In \fig{fig:pop-3a} we show that the population dynamics predicted by MRT show a markedly improved agreement with HEOM across the simulation set. While some deviations emerge at strong electronic couplings, MRT consistently captures the steady-state populations well.

\begin{figure}[ht]
    \centering
    \includegraphics[width=1.0\columnwidth]{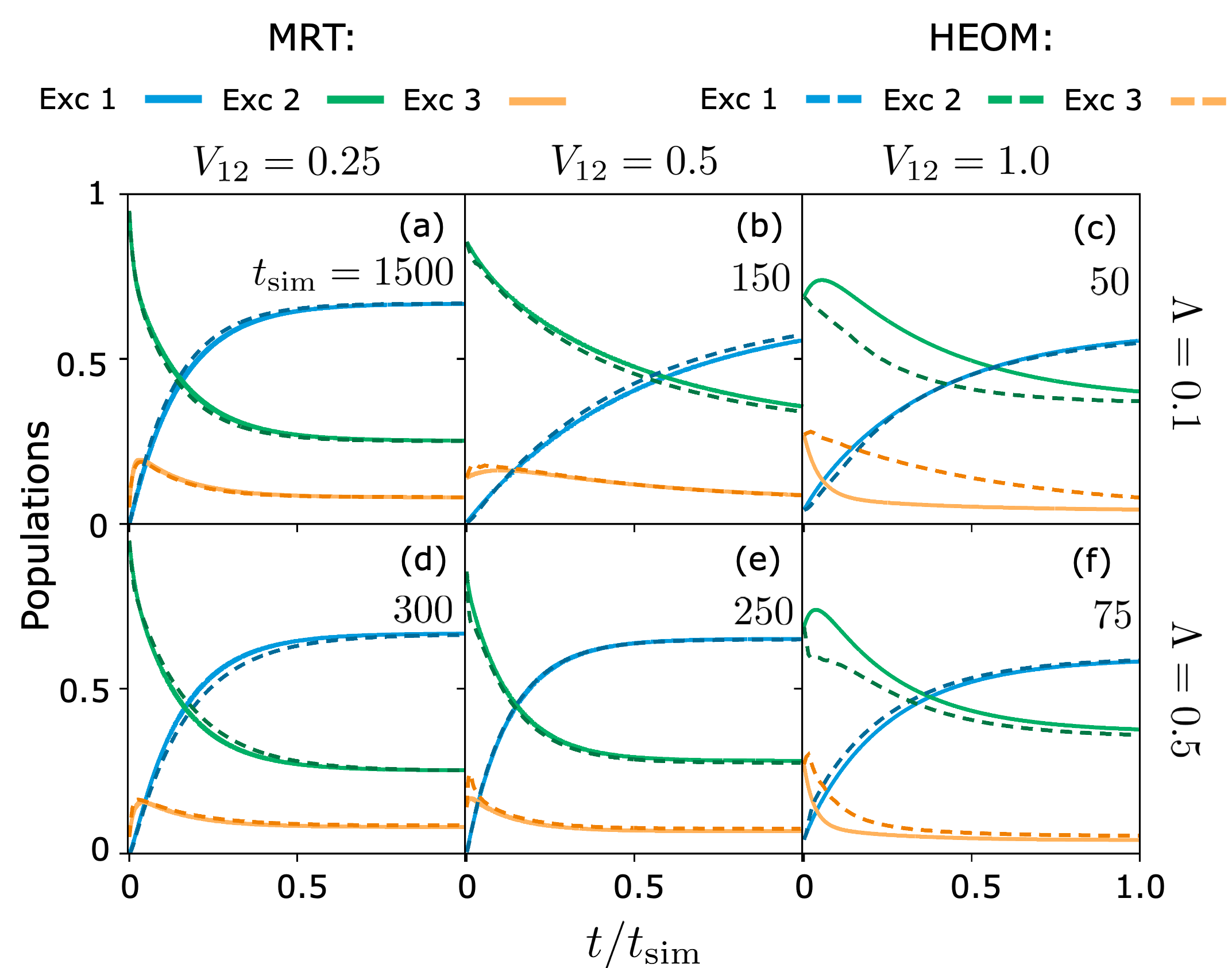}
    \caption{\textbf{Population dynamics for the molecular trimer (Simulation Set D) using an incoherent exciton initial condition corresponding to localization at site 2.}  Results from MRT calculations (full lines) are compared against HEOM (dashed lines). Exciton (Exc) populations are color-coded: Exc 1 (blue), Exc 2 (green), and Exc 3 (orange). The fixed simulation parameters are $T=1.0$, $\omega_{c}=0.5$, $\Delta E_{12}=1.0$, $\Delta E_{13}=2.0$,  $\Delta E_{23}=1.0$,  $V_{13}=0.25$ and $V_{23}=0.0$.}
    \label{fig:pop-3b}
\end{figure}

\begin{figure}[ht]
    \centering
    \includegraphics[width=1.0\columnwidth]{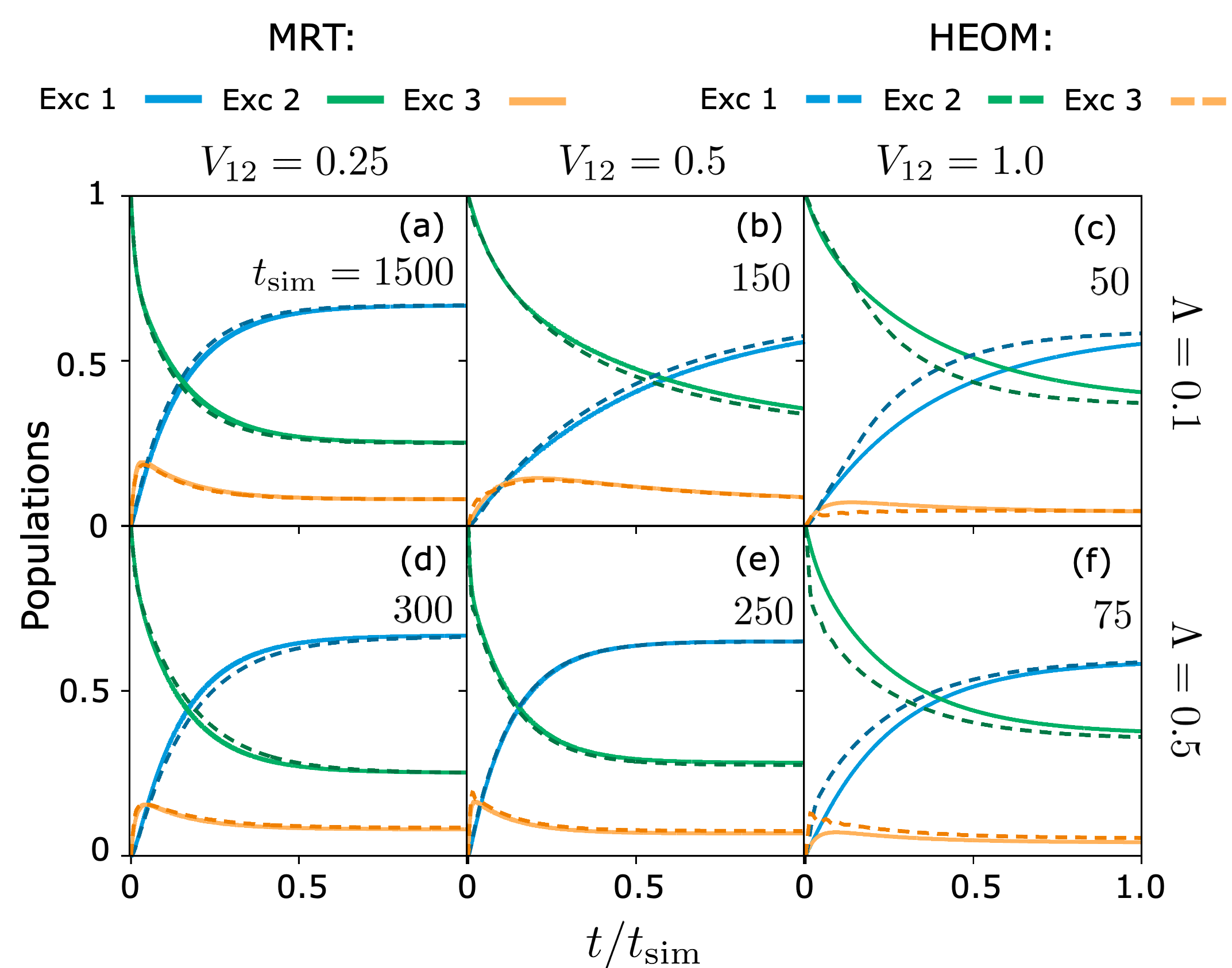}
    \caption{\textbf{Population dynamics for the molecular trimer (Simulation Set D) with the initial excitation localized in the middle exciton energy state.}  Results from MRT calculations (full lines) are compared against HEOM (dashed lines). Exciton (Exc) populations are color-coded: Exc 1 (blue), Exc 2 (green), and Exc 3 (orange). The fixed simulation parameters are $T=1.0$, $\omega_{c}=0.5$, $\Delta E_{12}=1.0$, $\Delta E_{13}=2.0$,  $\Delta E_{23}=1.0$,  $V_{13}=0.25$ and $V_{23}=0.0$. }
    \label{fig:pop-3c}
\end{figure}

Finally, building on the observation that MRT can effectively handle conditions challenging for QME, we tested an initial condition where the excitation is localized in the middle exciton energy state. The population dynamics for this scenario are shown in \fig{fig:pop-3c} (MRT: full lines, HEOM: dashed lines). The results again indicate good agreement between MRT and HEOM. This agreement persists until the electronic coupling becomes strong (e.g., $V_{12}=1.0$); however, even under such strong coupling, the steady-state populations are still accurately predicted by MRT.

\subsubsection{Site dissipation dynamics}

\begin{figure}[ht]
    \centering
    \includegraphics[width=1.0\columnwidth]{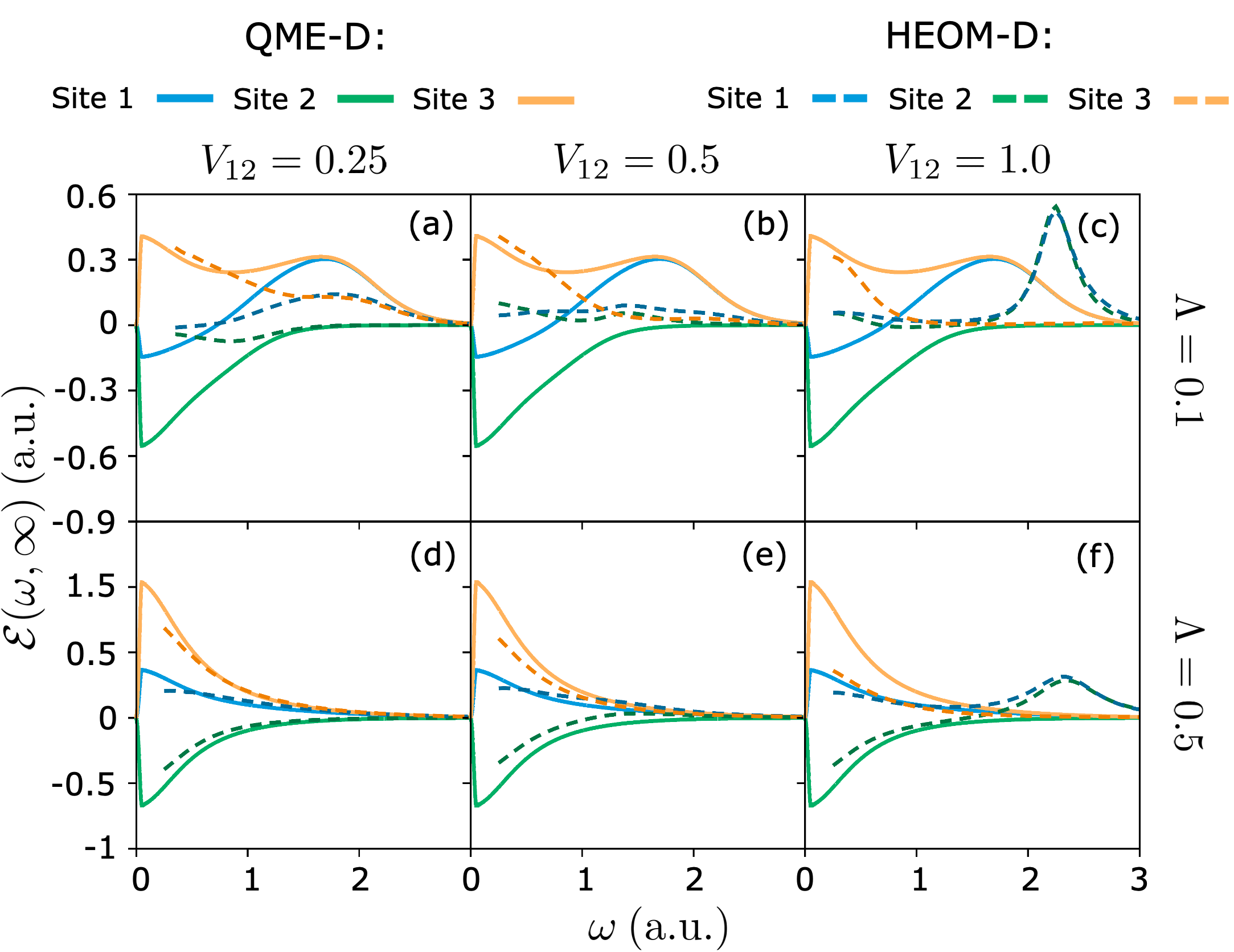}
    \caption{\textbf{Site steady-state dissipation density for the molecular trimer (Simulation Set D) with initial excitation localized at the middle site energy state.}  Results from QME-D calculations (full lines) are compared against HEOM-D (dashed lines). Site populations are color-coded: site 1 (blue), site 2 (green), and site 3 (orange). The fixed simulation parameters are $T=1.0$, $\omega_{c}=0.5$, $\Delta E_{12}=1.0$, $\Delta E_{13}=2.0$,  $\Delta E_{23}=1.0$,  $V_{13}=0.25$ and $V_{23}=0.0$.}
    \label{fig:dis-3a}
\end{figure}

We first focus on the QME-D method, with the system initialized with excitation localized at the middle site's energy state. The resulting steady-state site dissipation densities, $\mathcal{E}(\omega,\infty)$, are presented in \fig{fig:dis-3a}, comparing QME-D (full lines) with HEOM-D (dashed lines). The QME-D results align with the HEOM-D benchmarks only under specific conditions, notably at $V_{12}=0.25$ and $\Lambda=0.5$, \fig{fig:dis-3a}d. This agreement is expected as these parameters favor the perturbative treatment with respect to the electronic coupling $V$, a core assumption in the QME-D approach, see \stn{sec:theory-5}. The electronic coupling $V_{12}$ primarily scales the rate constants\cite{kim2024general1,kim2024general1} by $V_{12}^{2}$, which uniformly affects the entire frequency range of the dissipation spectrum. Consequently, while increasing $V_{12}$ speeds up the overall dissipation process, it does not significantly alter the qualitative features of the steady-state dissipation profiles within QME-D. However, for most other conditions, QME-D fails to accurately reproduce both the strength and the shape of the site dissipation densities compared to the HEOM-D benchmarks. Furthermore, HEOM-D reveals a structured peak at higher energies when $V_{12}=1.0$ (\fig{fig:dis-3a}c,f), consistent with the energy difference between the highest and lowest excitonic states. QME-D does not capture this feature.

\begin{figure}[ht]
    \centering
    \includegraphics[width=1.0\columnwidth]{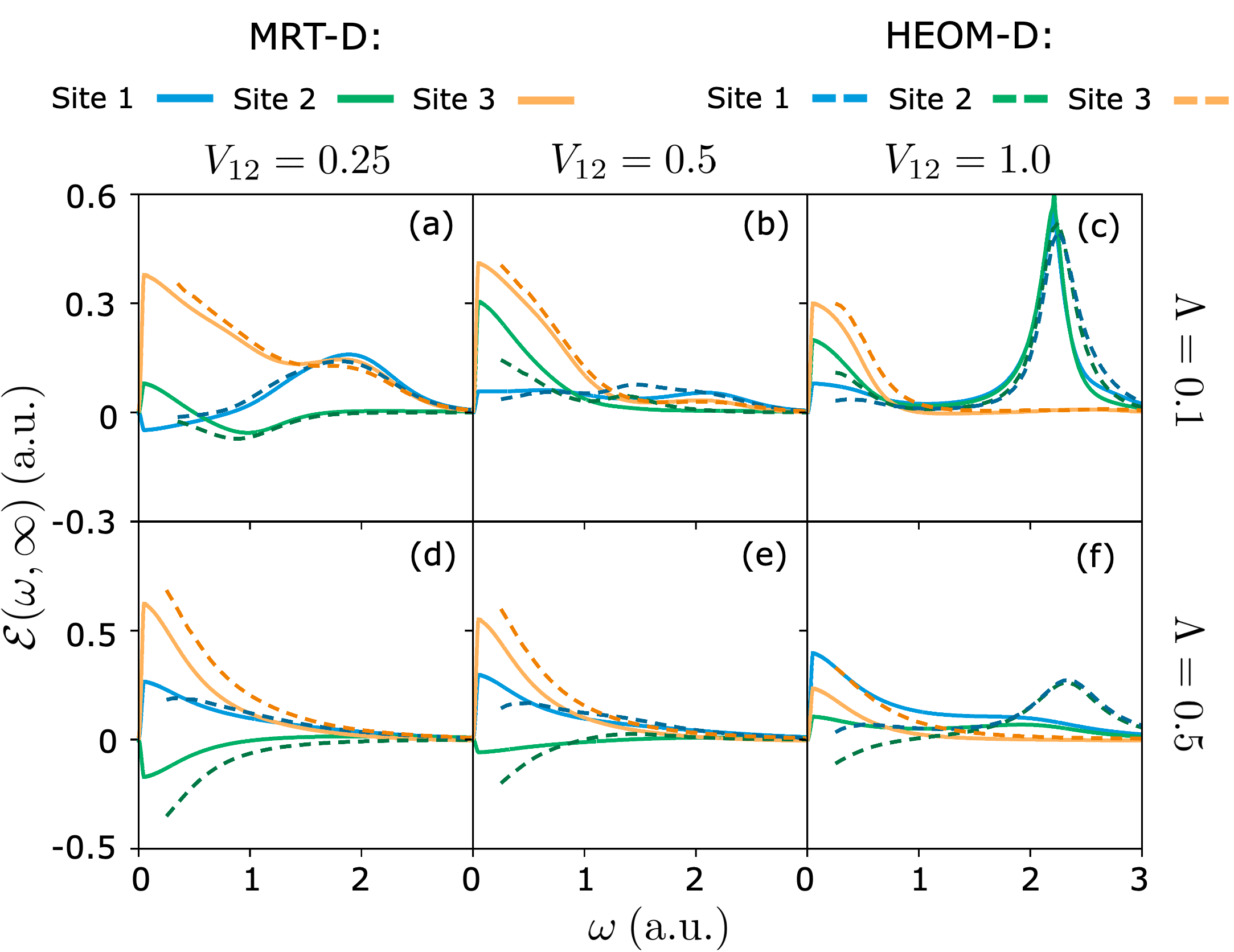}
    \caption{\textbf{Site steady-state dissipation density for the molecular trimer (Simulation Set D) using an incoherent exciton initial condition corresponding to localization at site 2.}  Results from MRT-D calculations (full lines) are compared against HEOM-D (dashed lines). Site populations are color-coded: site 1 (blue), site 2 (green), and site 3 (orange). The fixed simulation parameters are $T=1.0$, $\omega_{c}=0.5$, $\Delta E_{12}=1.0$, $\Delta E_{13}=2.0$,  $\Delta E_{23}=1.0$,  $V_{13}=0.25$ and $V_{23}=0.0$.}
    \label{fig:dis-3b}
\end{figure}

To address the limitations of QME-D, we now examine MRT-D. In \fig{fig:dis-3b} we show the site steady-state dissipation densities when the initial condition is an incoherent excitonic mixture corresponding to the localization at site 2. The HEOM-D results in \fig{fig:dis-3b} are very similar to those in \fig{fig:dis-3a}. Each site dissipation density predicted by MRT-D  closely follows the trends observed in the HEOM-D across a broader range of parameters, including the resonant structure of the peaks previously missed by QME-D for $V_{12}=1.0$. For $\Lambda=0.1$, MRT-D provides quantitatively accurate results (\fig{fig:dis-3b}a-c). Some discrepancies emerge as the reorganization energy increases to $\Lambda=0.5$ (\fig{fig:dis-3b}d-f). Nevertheless, MRT-D still captures the correct qualitative trends for $V_{12}=0.25$, \fig{fig:dis-3b}d, and $V_{12}=0.5$, \fig{fig:dis-3b}e, At strong coupling and high reorganization energy ($V_{12}=1$ and $\Lambda=0.5$), \fig{fig:dis-3b}f,  MRT-D struggle reproduce HEOM-D results. 

\begin{figure}[ht]
    \centering
\includegraphics[width=1.0\columnwidth]{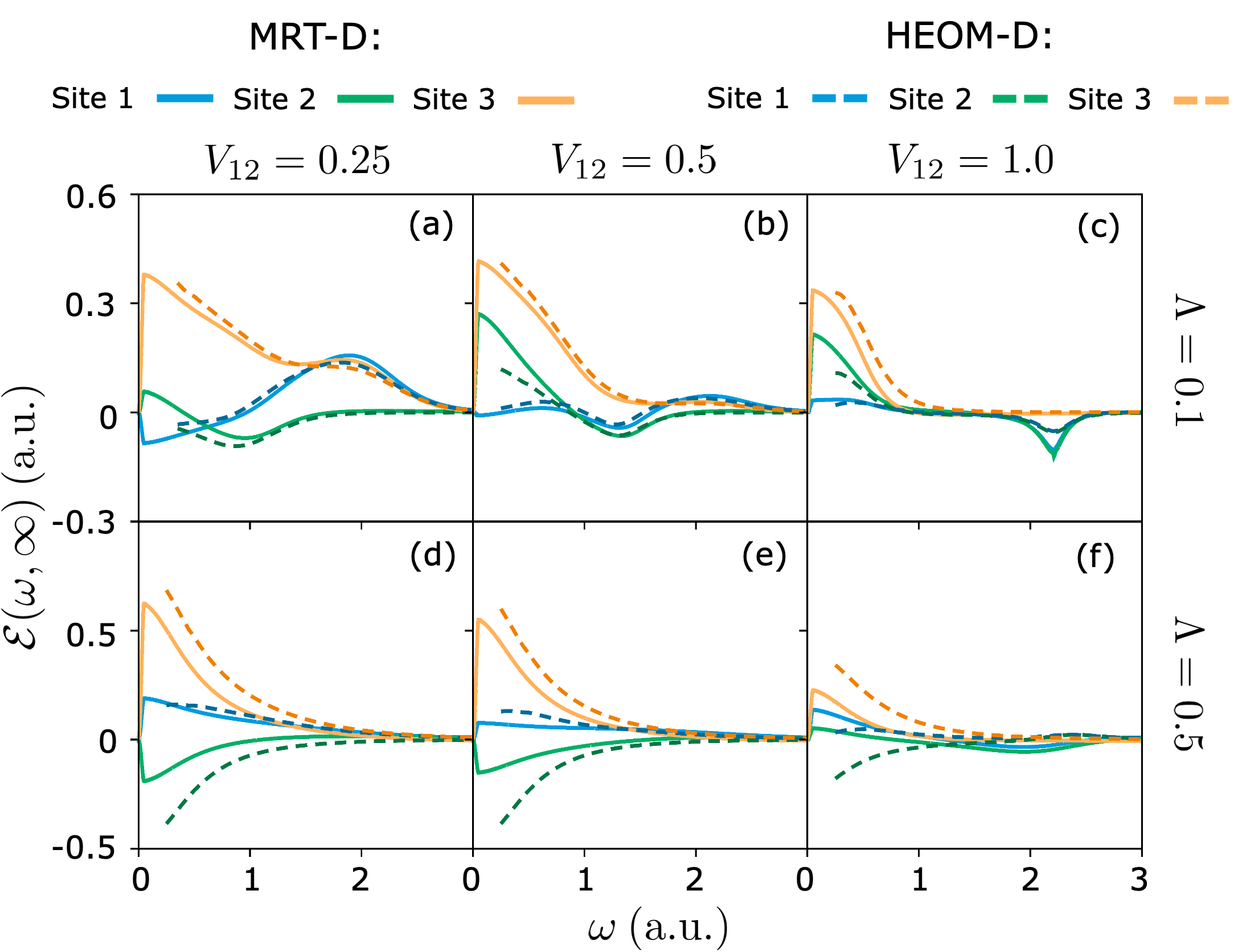}
    \caption{\textbf{Site steady-state dissipation density for the molecular trimer (Simulation Set D) with the initial excitation localized in the middle exciton energy state.}  Results from MRT-D calculations (full lines) are compared against HEOM-D (dashed lines). Site populations are color-coded: site 1 (blue), site 2 (green), and site 3 (orange). The fixed simulation parameters are $T=1.0$, $\omega_{c}=0.5$, $\Delta E_{12}=1.0$, $\Delta E_{13}=2.0$,  $\Delta E_{23}=1.0$,  $V_{13}=0.25$ and $V_{23}=0.0$.}
    \label{fig:dis-3c}
\end{figure}

Finally, we investigated the dissipation dynamics when the initial excitation is localized in the middle exciton energy state, with results shown in \fig{fig:dis-3c}. While the specific dissipation profiles differ from those in \fig{fig:dis-3b} (due to the different initial condition affecting the population dynamics that drive dissipation), MRT-D continues to effectively retrieve the correct trends and structural features of the dissipation dynamics across most conditions. The exception remains the challenging regime of strong electronic coupling and large reorganization energy ($V_{12}=1$ and $\Lambda=0.5$, \fig{fig:dis-3c}f), where deviations from HEOM-D are noticeable.

In summary, the MRT-D framework significantly extends and complements the QME-D approach, enabling the study of dissipation pathways across broader physical conditions. Importantly, while HEOM-D provides numerically exact results, its computational cost often limits its applicability to relatively small systems and baths with simple spectral densities. By contrast, approximate methods like QME-D and MRT-D offer a more scalable route to investigating dissipation in larger, more complex molecular systems with highly structured environments.

\section{Conclusions}\label{sec:conclusions-5}
In this paper, we have presented a significant advancement in understanding energy dissipation in open quantum systems by introducing a general theoretical framework that generalizes our previous theory of dissipation pathways in open quantum systems, QME-D, to include off-diagonal system-bath coupling mechanisms. This generalization is essential for more realistic treatment of molecular systems where such couplings are key in the quantum dynamics. Specifically, we provide a systematic derivation of quantum master equations describing population transfer and quantifying the energy dissipated into individual bath components. Furthermore, we provide rigorous proofs of energy conservation and detailed balance to establish the framework's physical integrity. 

The robustness and practical utility of the method were validated through its application to linearly coupled harmonic oscillator baths, which we referred to as MRT-D, as it is consistent with the Modified Redfield Theory. For this, we tested MRT-D against HEOM-D, a formally exact method. These tests included molecular dimers, spin-boson models, and molecular trimers, with baths described by Drude-Lorentz and Brownian oscillator spectral densities. Across a significant parameter range, MRT-D demonstrated good to excellent agreement with HEOM-D for both population dynamics and frequency-resolved dissipation spectra. Importantly, MRT-D successfully reproduced key spectral dissipation features, such as vibronic resonances in molecular dimers, site-dependent dissipation, and characteristic peaks in underdamped Brownian oscillator baths, highlighting the theory's ability to capture the detailed physics of the dissipation process. The advantages of MRT-D were particularly evident in molecular trimer simulations, where it resolves dissipation pathways in scenarios where QME-D fails. However, the validation studies of MRT-D also delineated its limitations, identifying regimes such as strong system-bath coupling, small energy gaps, or highly non-Markovian baths where its accuracy may be reduced.   

The application of our method is envisioned to be particularly powerful when integrated with sophisticated model Hamiltonians tailored for specific molecular systems. In particular, the method is well-suited to leveraging detailed spectral densities, whether derived from rigorous QM/MM simulations\cite{maity2021multiscale,kim2018excited,Cignoni2022Atomistic, jang2018delocalized,lee2016modeling,Kell2013shape,chen2023elucidating} or from fitting linearly coupled harmonic bath models to experimental spectroscopic data.\cite{ratsep2007electron,ratsep2008excitation,Pieper2011Excitonic,Pieper2009Chromophore,freiberg2009excitonic,gryliuk2014excitation,gustin_mapping_2023} The approach enables pinpointing the regions within these spectral densities that most significantly influence the system's dynamical evolution. Correlating these influential spectral features with molecular vibrational characteristics, in turn, can offer more comprehensive insights into how vibronic interactions steer non-adiabatic processes. These insights are helpful, for example, for accessing energy dissipation pathways through spectroscopic techniques such as 2D electronic spectroscopy.\cite{wit_extracting_2025} Ultimately, we envision a broad application of the framework to deepen our understanding of quantum dynamics across a wide range of complex molecular assemblies, including photosynthetic complexes\cite{mirkovic2017light,blau2018local,womick2011vibronic}, artificial excitonic\cite{hase2020designing,bolzonello_correlated_2016, yang_theoretical_2020, bialas_holsteinpeierls_2022} and plasmonic systems\cite{Hsu2017,bai_evolutionary_2021}, and molecular or solid-state qubits.\cite{gertler_protecting_2021,harrington_engineered_2022,chiesa_chirality-induced_2023}

\begin{acknowledgments}
C. W. K. was financially supported by the National Research Foundation of Korea (NRF) grants funded by the Ministry of Science and ICT of Korea (Grant Nos. 2022R1F1A1074027 and 2023M3K5A1094813) and Global - Learning $\&$ Academic research institution for Masters, PhD students, and Postdocs (LAMP) Program funded by Ministry of Education of Korea (Grant No. RS-2024-00442775). This material is based upon work supported by the U.S. Department of Energy, Office of Science, Office of Basic Energy Sciences, Quantum Information Science Research in Chemical Sciences, Geosciences, and Biosciences Program under Award Number DE-SC0025334.
\end{acknowledgments}

\renewcommand{\theequation}{A.\arabic{equation}}
\renewcommand{\thefigure}{A\arabic{figure}}
\renewcommand{\thetable}{A\arabic{table}}
\counterwithin*{equation}{section}
\section{Appendix}\label{appendix_5}

\subsection{Derivation of \eq{eq:tr123_ana}}\label{sec:app_traces}

This Appendix presents the detailed procedure for deriving \eq{eq:tr123_ana}, which are the analytical expressions of the traces in \eqs{eq:traces-b}{eq:traces-d} under a linearly coupled harmonic oscillator bath. For $\ti{Tr}1_{\beta\alpha}^{j}$ and $\ti{Tr}2_{\beta\alpha}^{j}$, we observe that 
\begin{equation} \label{eq:A1-chp5}
\begin{split}
        \frac{d{\tr}0_{\beta\alpha}^j}{dt'} &= - \frac{i}{\hbar} \tr_j \big[ (\hat{u}_\alpha^j)^\dagger (\hat{v}_{\beta\beta}^j - \hat{v}_{\alpha\alpha}^j) \hat{u}_\beta^j \hat{r}_\alpha^j \big] \\&=  - \frac{i}{\hbar} \tr_j \big[ (\hat{u}_\alpha^j)^\dagger \hat{u}_\beta^j (\hat{v}_{\beta\beta}^j - \hat{v}_{\alpha\alpha}^j) \hat{r}_\alpha^j \big]
\end{split}
\end{equation}
by expanding the bracket in the trace and recognizing the commutativity between the operators. We then plug in the expressions for $\hat{v}_{\alpha\alpha}^j$, $\hat{v}_{\beta\beta}^j$ and ${\tr}0_{\beta\alpha}^j$ [Eqs. (\ref{eq:vop_exci-a}), (\ref{eq:vop_exci-b}), (\ref{eq:tr0_ana})], and rearrange the resulting equations to yield
\begin{equation}\label{eq:A2-chp5}
\begin{split}
    \ti{Tr}_{j} [ (\hat{u}_{\alpha}^{j})^{\dagger}\hat{y}_{j}\hat{u}_{\beta}^{j}r_{\alpha}^{j}  ] &= \ti{Tr}_{j} [ (\hat{u}_{\alpha}^{j})^{\dagger}\hat{u}_{\beta}^{j}\hat{y}_{j}r_{\alpha}^{j}  ] \\&= \bigg( \frac{d_{\beta\beta}^{j}-d_{\alpha\alpha}^{j}}{2}[1+i\dot{f}(\omega_{j},t)]\bigg) \tr0_{\beta\alpha}^j.
\end{split}
\end{equation}
We can now insert \eqs{eq:vop_exci-c}{eq:vop_exci-d} into \eqs{eq:traces-b}{eq:traces-c}, respectively, and use the above results to arrive at \eqs{eq:tr123_ana-a}{eq:tr123_ana-b}.

For ${\tr}3_{\beta\alpha}^j$, we take the time derivative of \eq{eq:A1-chp5} to get
\begin{equation}
    \frac{d^2\ti{Tr}0_{\beta\alpha}^{j}}{dt'^{2}} = -\frac{1}{\hbar^{2}} \ti{Tr}_{j}[ (\hat{u}_{\alpha}^{j})^{\dagger} (\hat{v}_{\beta\beta}^j - \hat{v}_{\alpha\alpha}^j)\hat{u}_{\beta}^{j}(\hat{v}_{\beta\beta}^j - \hat{v}_{\alpha\alpha}^j)\hat{r}_{\alpha}^{j} ],
\end{equation}
from which we can derive
\begin{equation}
\begin{split}
        \ti{Tr}_{j}[(\hat{u}_{\alpha}^{j})^{\dagger}\hat{y}_{j}\hat{u}_{\beta}^{j}\hat{y}_{j}\hat{r}_{\alpha}^{j}]=&\bigg( \frac{(d_{\beta\beta}^{j}-d_{\alpha\alpha}^{j})^{2}}{4}[1+i\dot{f}(\omega_{j},t)]^{2}\\&+\frac{\hbar}{2\omega_{j}^{2}}\ddot{f}(\omega_{j},t)\bigg) \tr0_{\beta\alpha}^j
\end{split}
\end{equation}
with the help of \eq{eq:A2-chp5}. \Eq{eq:tr123_ana-c} then emerges from \eq{eq:traces-d} by taking a similar procedure as we did for $\tr1_{\beta\alpha}^j$ and $\tr2_{\beta\alpha}^j$.

\subsection{Discretization of the bath spectral densities}\label{sec:app_discretization} 

For the Drude-Lorentz (DL) spectral density [\eq{eq:DL-SPD-5}], our discretization scheme follows Ref. \cite{wang_semiclassical_1999}. Individual bath modes are positioned at frequencies $\omega_{j}$ according to:
\begin{equation}\label{eq:discrete-w}
    \omega_{j}=\frac{j^{2}}{N^{2}}\omega_{\text{max}}, \:\:\: \text{for}\:\: j=1,2,...,N.
\end{equation}
Here, $N$ is the total number of discrete modes representing the BSD, and $\omega_{\text{max}}$ is a defined upper frequency cutoff. This formula results in a denser distribution of modes at lower frequencies, which is appropriate because the reorganization energy density, $J_{\text{DL}}(\omega)/\omega$, is typically larger in this region.

We now define the function $f_{\text{DL}}(\omega)$, which links the discrete and continuous representation of the spectral density as
\begin{equation}
    \frac{\omega_{j}^{3}d_{j}^{2}}{2}=\frac{J_{\text{DL}}(\omega_{j})}{f_{\text{DL}}(\omega_{j})}.
\end{equation}
The explicit form of this function is given by
\begin{equation}\label{eq:f-w}
    f_{\text{DL}}(\omega)=\frac{N}{2\sqrt{\omega\omega_{\text{max}}}}
\end{equation}
which allows us to obtain the discrete reorganization energy, $\lambda_{j}$, as
\begin{equation}
    \lambda_{j}=\frac{\omega_{j}^{2}d_{j}^{2}}{2}=\frac{4\Lambda}{j\pi}\frac{\omega_{c}\omega_{j}}{\omega_{j}^{2}+\omega_{c}^{2}}.
\end{equation}
This discrete $\lambda_{j}$ is consistent with the reorganization energy obtained by integrating the continuous $J_{\text{DL}}(\omega)/\omega$ function over a frequency segment corresponding to the $j$-th mode
\begin{equation}
        \int_{\text{segment } j} \frac{J_{\text{DL}}(\omega)}{\omega} d\omega \approx \frac{J_{\text{DL}}(\omega_{j})}{\omega_{j}} \Delta\omega_{j}=\frac{4\Lambda}{j\pi}\frac{\omega_{c}\omega_{j}}{\omega_{j}^{2}+\omega_{c}^{2}}
\end{equation}
where $\Delta \omega_{j}=[(\omega_{j}+\omega_{j+1})/2]-(\omega_{j}+\omega_{j-1})/2]$.

For the Brownian Oscillator (BO) spectral density [\eq{eq:BO-SPD}], assuming a cutoff $\omega_{\text{max}}>\omega_{0}$ (the characteristic frequency of the BO), the first step is to find the frequency $\Omega$ that maximizes the reorganization energy density $J_{\text{BO}}(\omega)/\omega$ within the range $[0,\omega_{\text{max}}]$
\begin{equation}
    \Omega = \sqrt{\text{max}[0,\omega_{0}^{2}-2\gamma^{2}]}
\end{equation}
where $\gamma$ is the damping parameter.
If $\Omega=0$, the discretization strategy is similar to that for the DL spectral density, using Eqs. \ref{eq:discrete-w}-\ref{eq:f-w}, but substituting $J_{\text{BO}}(\omega)$ for $J_{\text{DL}}(\omega)$.
Otherwise, the frequency domain is split into two windows, $[0,\Omega)$ and $(\Omega,\omega_{\text{max}}]$. Each window is represented by $N/2$ bath modes using different discretization schemes.
For the $[0,\Omega)$ window
\begin{equation}
\begin{split}
        \omega_{1,j}&=\left[1-\left(1-\frac{2j}{N}\right)^{2}\right]\Omega, \\ f_{BO1}(\omega)&=\frac{N}{4\sqrt{(\Omega-\omega)\Omega}}, \quad \text{for } j=1,2,...,\frac{N}{2}-1 
\end{split}
\end{equation}
For the $(\Omega,\omega_{\text{max}}]$ window 
\begin{equation}
\begin{split}
        \omega_{2,j}&=\Omega+\frac{4j^{2}}{N^{2}}(\omega_{max}-\Omega), \\ f_{BO2}(\omega)&=\frac{N}{4\sqrt{(\omega-\Omega)(\omega_{max}-\Omega)}}, \quad \text{for } j=1,2,...,\frac{N}{2}.
\end{split}
\end{equation}
While this discretization does not include a mode exactly $\omega=\Omega$, where both $f_{BO1}(\omega)$ and $f_{BO2}(\omega)$ diverge, we can fix this by assigning a specific reorganization energy, $\lambda_{\omega=\Omega}$ to a mode placed at $\omega=\Omega$. This value is determined by ensuring that the reorganization energy of this discrete mode matches the integrated reorganization energy from the continuous $J_{\text{BO}}(\omega)$ over a small frequency interval centered at ${\Omega}$, which yields
\begin{equation}
    \lambda_{\omega=\Omega}=\frac{2\Lambda}{\pi N^{2}}\frac{\omega_{\text{max}}\omega_{0}^{2}}{\gamma(\omega_{0}^{2}-\gamma^{2})}
\end{equation}
This discretization scheme ensures that the bath modes are concentrated around $\omega=\Omega$, the region with most significant system-bath coupling, analogous to the low-frequency concentration for the DL density.

\bibliography{references}

\end{document}